\documentclass[%
 aip,
 amsmath,amssymb,
 reprint,%
]{revtex4-1}

\usepackage{graphicx}% Include figure files
\usepackage{dcolumn}% Align table columns on decimal point
\usepackage{bm}% bold math
\usepackage[utf8]{inputenc}
\usepackage[T1]{fontenc}
\usepackage{mathptmx}

\usepackage{amsmath,amssymb}
\usepackage{graphicx}
\usepackage{verbatim}
\usepackage{enumitem}
\usepackage{color}
\usepackage[english]{babel}
\usepackage{bm}
\usepackage{natbib}
\usepackage[symbol]{footmisc}
\usepackage{textcomp}

\DeclareUnicodeCharacter{2212}{-}
\DeclareUnicodeCharacter{308}{}
\DeclareUnicodeCharacter{301}{}

\begin{document}

\preprint{}

\title[]{Active meta-optics and nanophotonics with halide perovskites}

\author{Alexander S. Berestennikov}
 \affiliation{ITMO University, St.~Petersburg 197101, Russia}
 \author{Pavel M. Voroshilov}
 \affiliation{ITMO University, St.~Petersburg 197101, Russia}
\author{Sergey V. Makarov}
\affiliation{ITMO University, St.~Petersburg 197101, Russia}
\author{Yuri S. Kivshar}
\affiliation{ITMO University, St.~Petersburg 197101, Russia}
\affiliation{Nonlinear Physics Center, Australian National University, Canberra ACT 2601, Australia}

\date{\today}% It is always \today, today,

\begin{abstract}
Meta-optics based on optically resonant all-dielectric structures is a rapidly developing research area driven by its potential applications for low-loss efficient metadevices.  Active, light-emitting subwavelengh nanostructures and metasurfaces are of a  particular interest for meta-optics, as they offer unique opportunities for novel types of compact light sources and nanolasers. Recently, the study of {\it halide perovskites} has attracted an enormous attention due to their exceptional optical and electrical properties. As a result, this family of materials can provide a prospective platform for modern nanophotonics and meta-optics, allowing to overcome many obstacles associated with the use of conventional semiconductor materials. Here we review the recent progress in the field of halide-perovskite meta-optics with the central focus on light-emitting nanoantennas and metasurfaces for the emerging field of {\it active metadevices}. 
\end{abstract}

\maketitle

\section{\label{sec:level1}Introduction}

\begin{figure*}
\centering
\includegraphics[width=0.75\linewidth]{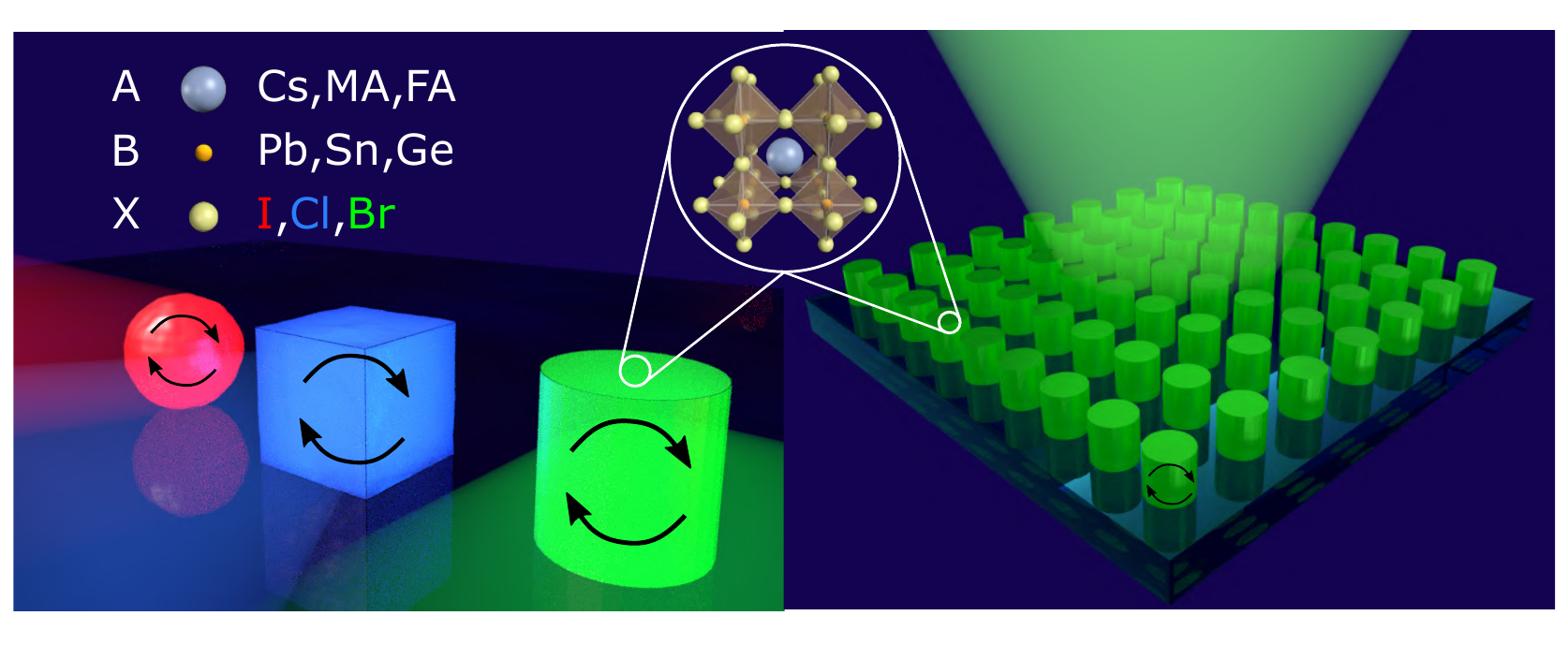}% Here is how to import EPS art
\caption{\label{fig:main} Schematic illustration of the basic designs of meta-optics structures based on halide perovskites for metadevice applications: Individual particles of various shapes and arrays of nanoparticles with unique emission properties tunable in the entire optical spectral range. Cations `MA' and `FA' stand for CH$_3$NH$_3^+$ and HC-(NH$_2$)$_2^+$, respectively.}
\end{figure*}

Nanophotonics bridges optics and nanoscale science, and it describes localization of light at the nanoscale in spatially trapped resonant optical modes~\cite{review_albert}, enabling to scale down substantially the size of many optical devices. For many years, the management of light in the subwavelength regime was associated solely with metallic structures supporting plasmonic resonances, paving the way to various nanoscale optical phenomena and applications of {\it nanoplasmonics}~\cite{plasmonics}.

To confine light at the nanoscale, traditional nanophotonics employs noble metals, as well as various metal alloys and doped oxides~\cite{review_albert,plasmonics}. Recently, to overcome optical losses and bring novel functionalities, optically resonant {\it dielectric nanostructures} have been introduced and extensively studied over the last decade~\cite{kuznetsov2016optically,kruk}. Usually, conventional semiconductors such as silicon or gallium arsenide are employed for such nanostructures due to their high values of the refractive index and well-developed fabrication methods~\cite{optica}. However, most semiconductors face some limitations related to difficulties with reversible spectral tunability, still expensive fabrication processes, and low quantum yield at room temperatures. Therefore, more complicated designs with integrated quantum dots or quantum wells had to be applied~\cite{staude2019} to overcome such limitations partially. We believe that the next important step is the simplification of the developed designs making them more attractive for large-scale low-cost technological applications by utilizing a rising star of the modern material science – {\it halide perovskites}.

Halide perovskites are materials with a composition ABX$_3$, where X stands for halide (I$^-$, Br$^-$ or Cl$^-$), A is an organic or inorganic element, and B is one of the cations listed in Fig.~\ref{fig:main}. They support excitonic states at room temperature, refractive indices ($n=$2$\div$3) high enough for the efficient excitation of various optical resonances, chemically tunable bandgap, strong optical nonlinear response, high defect tolerance, and, thus, quantum yield in the range 30$\div$90\% of photoluminescence (PL) at room temperatures~\cite{kim2016metal, sutherland2016perovskite}. They have already revolutionized the solar-cell technologies with up to 24.2\% performance efficiency~\cite{NREL}. Next breakthrough in this field will lead to a novel generation of perovskite-based light-emitting diodes~\cite{lin2018perovskite} and photodetectors, and the efficiencies of such devices grow rapidly with a promise to overtake many previously developed technologies. In turn, light management at the nanoscale is highly important for such a kind of applications, because light absorption and emission occurs at the subwavelength scale. This raises an important question: {\it Can nanophotonics make perovskite-based optical devices substantially better?} The recent results suggest that indeed perovskites can be empowered by optical resonances and nanostructuring, as illustrated schematically in Fig.~\ref{fig:main}.

Importantly, tight confinement of the electromagnetic fields in resonant optical nanostructures can boost light-matter interaction, and thus enhance light emission processes. All-dielectric light-emitting (\textit{`active'}) nanostructures offer many advantages over their plasmonic counterparts, including high radiation efficiency, directional far-field coupling, and lower heat generation. Furthermore, light sources can be directly placed inside the nanoresonator material, where the near-field enhancement is usually the largest. Nanostructures with optical resonances at the wavelength of the material emission can offer many new opportunities for controlling light emission at the nanoscale, providing a pathway for the realization of novel types of light sources and nanolasers. 

In this paper, we review the current progress in this recently emerged field of halide-perovskite meta-optics driven by multipolar Mie-type resonances and summarize the most remarkable developments. In addition, we discuss some future research directions and expected novel applications of this field, including lasing from nanoparticles and metasurfaces, as well as chemically tunable light-emitting metadevices.

\section{What is "meta-optics" ?}

{\it "Meta-optics"} is a new and rapidly developing research direction in subwavelength optics and nanophotonics, and it is inspired by the physics of metamaterials~\cite{kruk} where the electromagnetic response is associated with the magnetic dipole resonances and optical magnetism originating from the resonant dielectric nanostructures with high refractive index~\cite{kruk}. The concepts of meta-optics and all-dielectric resonant nanophotonics are driven by the idea to employ subwavelength dielectric Mie-resonant nanoparticles as "meta-atoms" for creating highly efficient optical metasurfaces and metadevices, the latter are defined as devices having unique functionalities derived from structuring of functional matter on the subwavelength scale~\cite{zheludev_nmat}. 

To the best of our knowledge, the term "meta-optics" associated with this context was suggested by Professor Emeritus Ross McPhedran~\cite{ross} who commented on this as follows: {\it “I think I made the term up to denote traditional optics but enriched with new elements provided by metamaterials”} (Ross McPhedran, University of Sydney: March 5, 2018). Indeed, the term  "meta-optics" emphasizes the importance of optically-induced magnetic response of artificial subwavelength-patterned structures. Because of the unique optical resonances and their various combinations employed for interference effects accompanied by strong localization of the electromagnetic fields, high-index nanoscale structures are expected to complement or even replace different plasmonic components in a range of potential applications. Moreover, many concepts which had been developed for plasmonic structures, but fell short of their potential due to strong losses of metals at optical frequencies, can now be realized based on Mie-resonant dielectric structures.

\section{Halide perovskites vs. semiconductors}

During last few years monolithic light-emitting nanoantennas and metasurfaces have been fabricated from various semiconductor materials~\cite{staude2019all}. PL enhancement was observed from Si nanocrystals incorporated in SiO$_2$ resonant nanoparticles~\cite{capretti2017integrating}, Ge quantum dots in Si resonant nanoparticles~\cite{rutckaia2017quantum}, and NV-centers in resonant diamond nanoparticles~\cite{zalogina2018purcell}. Aiming for more general applications, we discuss the advantages of halide perovskites and compare them with conventional semiconductors such as Si and GaAs, because they are widely used in modern nanophotonics~\cite{staude2017metamaterial, xiang2017hot,liu2018light, ha2018directional}, as well as optoelectronics and photovoltaics. 

\begin{figure*}
\centering
\includegraphics[width=0.8\linewidth]{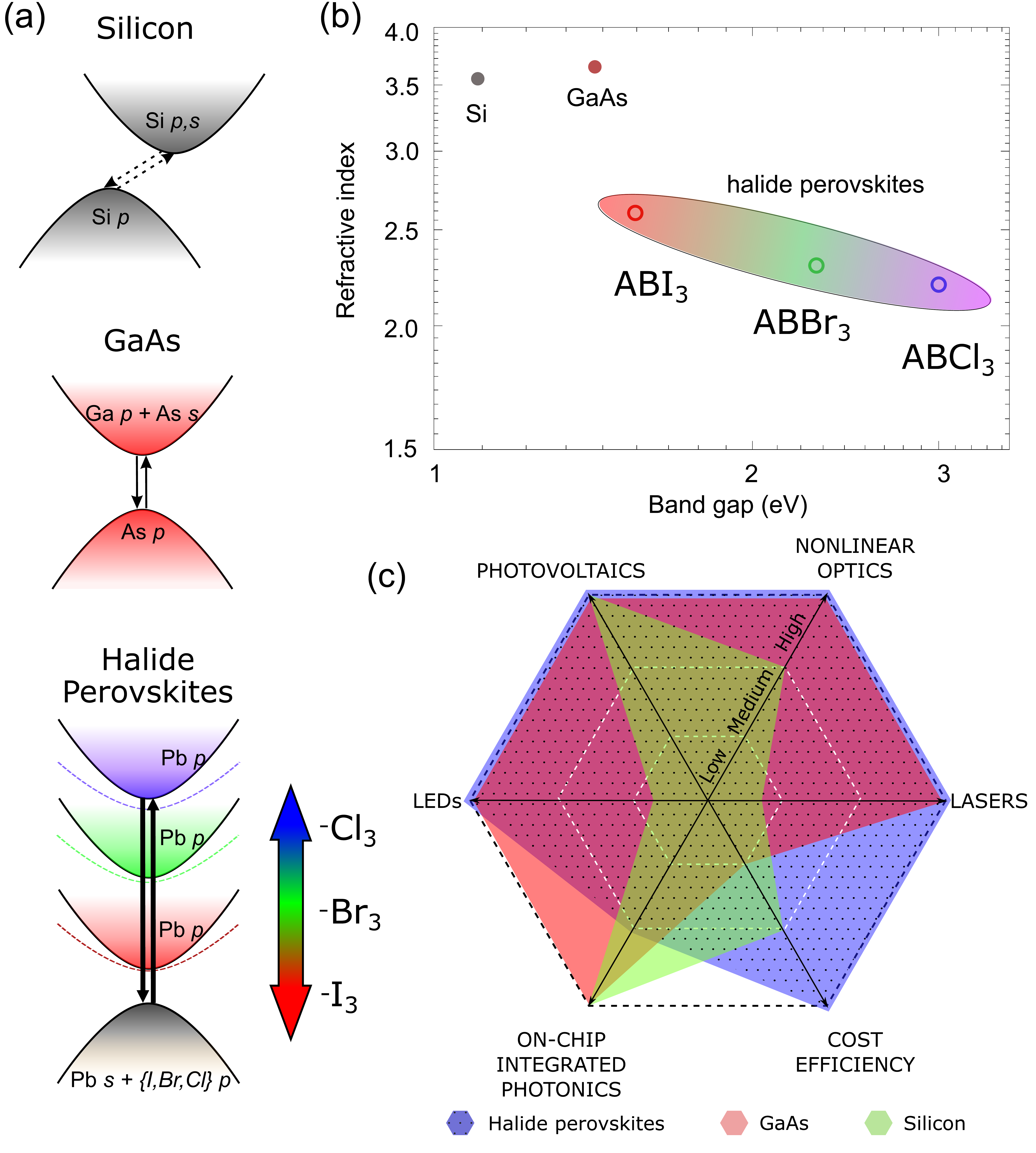}
\caption{\label{fig:properties} Comparison of conventional semiconductor materials and halide perovskites. (a) Sketches of band structures for different semiconductors (Si, GaAs, and halide perovskites), indicating broadband and reversible tuning of band gap for perovskites via anions variation (a rainbow arrow). Dashed parabolic curves illustrate the existence of excitons in the perovskites at room temperature. (b) Real parts of refractive index (\textit{n}) around the edges of conduction band are shown for Si, GaAs, and halide perovskites, where A stands for Cs,MA, or FA, and B stands for Pb, Sn, or Ge. Hollow circles correspond to MAPbX$_3$ compositions. (c) A hexagon scheme depicts comparison of suitability of various materials (Si, GaAs, and halide perovskites) for different applications. Low level corresponds to strong limitations for material to be applied in a particular field; medium level implies the moderate limitations; high level means that the material either is already widely used or highly prospective for a particular field of applications.}
\end{figure*}

As shown schematically in Fig.~\ref{fig:properties}a, halide perovskites can be considered as direct band-gap materials resembling GaAs. Ab initio calculations~\cite{yin2015halide} show that the state contributed by the cation A (e.g. Cs, MA, FA, etc.) is far from the band edges, which means that it does not play a significant role in determining the basic electronic structures in halide perovskites. In turn, the valence band has Pb \textit{s} and X \textit{p} anti-bonding character, whereas the conduction band is formed by the Pb \textit{p} states. This is unique dual nature (ionic and covalent) of halide perovskites electronic structure. In conventional semiconductors such as GaAs, the conduction band primarily has \textit{s} orbital character, whereas the valence band primarily has \textit{p} orbital character. In contrast, halide perovskites exhibit inverted band structure, and the conduction band is mainly composed of degenerate Pb \textit{p} bands. The atomic \textit{p} orbital are less dispersive than \textit{s} orbitals, and the density of states in the lower conduction band of the halide perovskites is significantly higher than in GaAs, and it is several orders of magnitude stronger than in Si. As a result, halide perovskites have stronger the interband transition dipole moment, yielding a high absorption coefficient and high quantum yield for the emission of light. 

Additionally, as compared to GaAs, halide perovskites also have high tolerance to defects for luminescence~\cite{kang2017high}, excitons at room temperature~\cite{tanaka2003comparative}, slightly lower refractive index in visible/near-infrared ranges, and reversibly tunable band gap from 1.5~eV up to 3~eV~\cite{pellet2015transforming, solis2015post} (Fig.~\ref{fig:properties}b). Due to these favorable characteristics, ease of processing and low price, the perovskites are the promising family of material for the next generation optoelectronic and nanophotonic devices.  Namely, there is an urgent necessity to create cheap, efficient, reliable and easy-to-manufacture solar cells, LEDs, ultracompact lasers, nonlinear optical devices, and optical chips. Another scheme in Fig.~\ref{fig:properties}c shows general comparison of applications of halide perovskites with those for Si and GaAs based technologies, where meta-optics and currently developing the concepts can boost the performance of real devices. 

\subsection{On-chip integrated photonics}

The real part of the refractive index of halide perovskites is significantly larger than that for SiO$_2$ or most of polymers, making the perovskites to be good materials for on-chip integration, owing to high enough optical contrast. Moreover, \textit{in situ} chemical variation of the perovskites band gap~\cite{pellet2015transforming, solis2015post} and, thus, the spectrum of their luminescence, opens up unprecedented opportunities for reconfigurable nanophotonic devices not achievable with the use of the conventional GaAs platform. As a result, halide perovskite nanoparticles~\cite{tiguntseva2018light}, nanowires~\cite{zhu2015lead,xing2015vapor}, microdisks~\cite{cegielski2018monolithically}, microplates~\cite{zhang2014room}, nanoscale gratings~\cite{tiguntseva2019enhanced}, and metasurfaces~\cite{gholipour2017organometallic,makarov2017efficient} are easy to fabricate and process, and they can become a convenient and cheap part of optical circuitry in the near future~\cite{makarov2019halide}. Nevertheless, the existing silicon-based platform is much more developed for on-chip integrated photonics~\cite{bogaerts2018silicon}, but it is expected to be outperformed by the perovskites in the future.

\subsection{Nonlinear optics}

Nonlinear properties of halide perovskites are superior to many conventional semiconductors~\cite{ferrando2018toward, xu2019halide}. For example, CsPbX$_3$ demonstrates two-photon, three-photon and even five-photon absorption~\cite{chen2017giant}, being sufficient for multiphoton pumping of lasing~\cite{xu2016two}. Third-harmonic generation has also been measured with high third-order susceptibilities $\chi^{(3)}$ exceeding the level of  10$^{-17}$ m$^2$V$^{-2}$[~\cite{abdelwahab2017highly}]. Second-harmonic generation from halide perovskite (CsPbX$_3$ or MAPbX$_3$, where MA=CH$_3$NH$_3$) structures is not so effective due to the centrosymmetric crystalline structure. However, lead-free perovskites (such as CsGeI$_3$ and MAGeI$_3$) exhibit giant second-harmonic generation with the second-order nonlinear coefficient $\chi^{(2)}$ exceeding the values of 150~pm~V$^{-1}$, being comparable with those for GaAs and exceeding by many orders the coefficients observed for silicon~\cite{cazzanelli2012second, makarov2017efficient}.

\subsection{LEDs and lasers}

Because of a direct bandgap, strong excitons, high defect tolerance, and outstanding color tunability of halide perovskites, they are used actively for light emitting devices (LEDs)~\cite{tan2014bright}. Best performing  perovskite-based LEDs demonstrate the efficiencies higher than 20\%~\cite{lin2018perovskite, xu2019rational}, and this number can be improved further by smart nanophotonic designs~\cite{makarov2019halide}, as was demonstrated for other LED materials~\cite{wiesmann2009photonic}.

Miniaturization of light-emitting architectures is motivated by the development of nanolasers and microlasers~\cite{ma2019applications}. Perovskite lasers based on nanowires, microdisks, and nanoscale gratings show low-lasing thresholds, spectral tunability, and high defect stability~\cite{zhu2015lead, veldhuis2016perovskite, sutherland2016perovskite, makarov2019halide}.
In turn, pure Si and GaAs can not compete with the perovskite lasers, and they need an additional doping or integration with quantum wells. However, Si- and GaAs-based sources emitting in the near IR frequency range are currently widely used in telecom applications, where no compositions for halide perovskite are available yet. Also, Si based nanostructures can emit white light upon strong photoexcitation with relatively low quantum yield~\cite{makarov2017nanoscale, zhang2018lighting}, which can be outperformed by layered rather than bulk perovskites~\cite{smith2019tuning}.

\subsection{Photovoltaics}

Most of commercial photovoltaic solar cells have been produced using silicon. Wafer-based monocrystalline or polycrystalline silicon solar cells exhibit stability, heat resistance, relatively high photo-conversion efficiency (up to $\approx$26\%) and low-cost fabrication. However, the main disadvantage of crystalline silicon is a poor absorbance due to indirect bandgap, which requires an increase of an active layer. It makes problematic creating flexible thin-film photovoltaic devices. The use of GaAs allows to avoid this disadvantage, and it is well suited for photoelectric applications, with an optimal direct bandgap (1.424 eV), high absorption coefficient, good carrier lifetime and mobility, as well as low nonradiative energy losses. A key disadvantage is a high cost of production due to the need to use the epitaxial growth methods. 

Iodine perovskites (e.g. MAPbI$_3$) have the optical characteristics close to GaAs, while their electronic properties make the perovskites favorable for thin-film photovoltaics owing to stronger light absorption~\cite{green2014emergence, yin2015halide}. At the same time, halide perovskite are based on earth-abundant and cheap materials, as well as easy processing without complex manufacturing multistage procedures. Remarkably, by employing the concepts of nanophotonics and meta-optics, one can improve further the characteristics of perovskite solar cells~\cite{zhang2013enhancement, furasova2018resonant, jimenez2018absorption, zhang2019photonics}.

\subsection{Cost efficiency}

Due to abundance of silicon, streamlined production, and high density of components which can be placed on a substrate, the cost of silicon components is very low. However, complex methods of creating and post-processing of silicon-based structures are still required for the fabrication of silicon micro- and nanostructures. Devices based on GaAs are several orders of magnitude more expensive due to a high cost and small size of the substrates, as well as the use of complex epitaxial growth methods and rarity of the material. In this category, perovskite-based devices look more affordable and cheaper for the production on large scales~\cite{snaith2018present}. Using wet chemistry methods and roll-to-roll techniques, it is possible to create flexible, cheap, and effective optoelectronic and nanophotonic devices. Finally, perovskite based devices have also now been demonstrated to be stable for thousands of hours under conditions that should mimic those that a real device can be exposed to.

\begin{figure}[h!]
\centering
\includegraphics[width=0.95\linewidth]{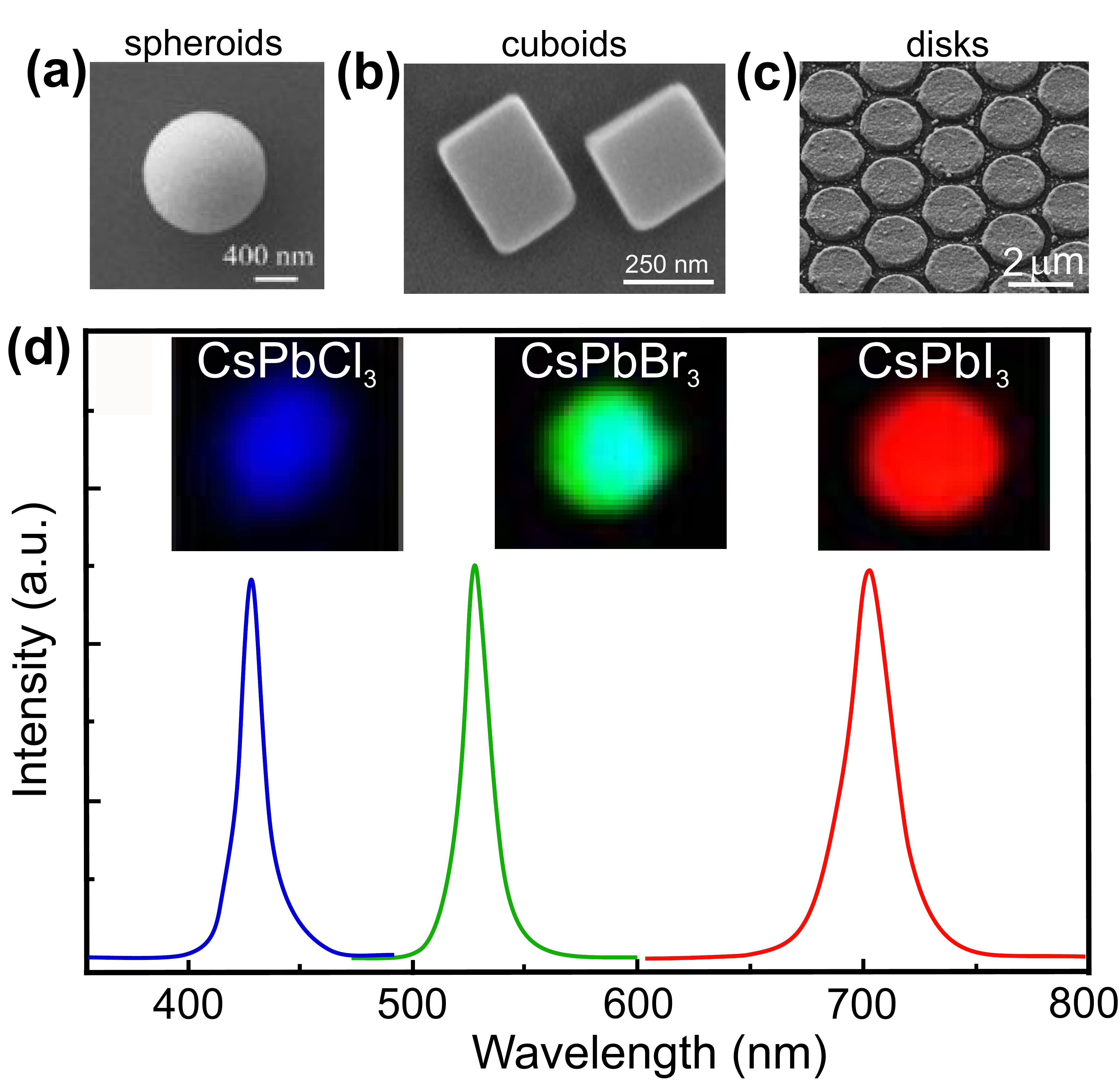}% Here is how to import EPS art
\caption{\label{fig:Blocks} SEM images of major building blocks of halide perovskite meta-optics: (a) spheroids,~\cite{tang2017single} (b)  cuboids,~\cite{liu2018robust} and (c)  disks.~\cite{zhizhchenko2019single} (d) Photoluminescence spectra for perovskite spheroids with various compositions. Adapted with permission from~\cite{tang2017single}.}
\end{figure}

\begin{figure*}
\centering
\includegraphics[width=0.99\linewidth]{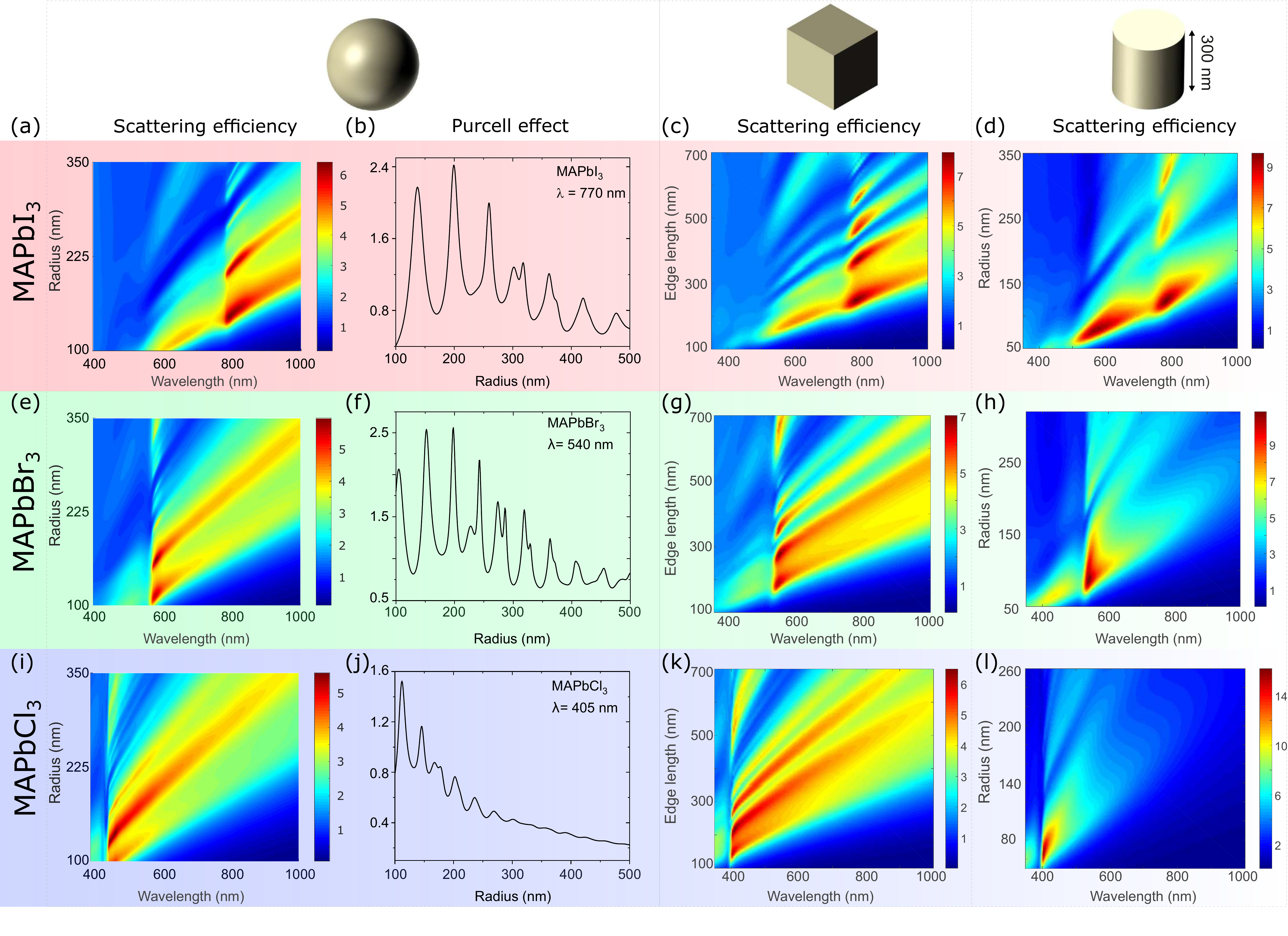}% Here is how to import EPS art
\caption{\label{fig:Mie} Scattering properties of halide-perovskite nanoparticles of different shapes. Calculated scattering efficiencies for spheres, cubes, and cylinders, and the average Purcell factor for spherical particles. Results are presented for three typical chemical compositions of the halide perovskite materials: MAPbI$_3$, MAPbBr$_3$, and MAPbCl$_3$, where MA=CH$_3$NH$_3$.}
\end{figure*}

\section{Basic concepts and tools}

\subsection{Optical resonances in dielectric meta-optics}

Resonances play a crucial role in meta-optics because they allow to enhance both electric and magnetic fields. Unlike plasmonic nanoparticles and their structures, the electromagnetic fields penetrate into dielectric nanostructures, and the light-matter interaction in such structures depends strongly on geometric resonances and interferences between different multipolar modes for individual elements and their compositions such as oligomers and metasurfaces. 

A building block of any metamaterial or metasurface is a resonant subwavelength structure, frequently called {\it `meta-atom’} and also known in plasmonics as `nanoantenna’. Recently, optically-resonant dielectric nanoparticles with high refractive index have attracted a lot of attention in nanophotonics as an alternative approach to achieve a strong Mie-resonant response accompanied by subwavelength localization of light~\cite{kuznetsov2016optically}. Meta-optics in halide perovskites takes an advantage of their relatively high refractive index and a possibility to employ the structures with different shapes. Typical shapes of halide-perovskite meta-atoms are spheres, cubes, and cylinders, as shown in Figs.~\ref{fig:Blocks}a-c. Optical response of dielectric nanoparticles of various shapes has been studied intensively during last few years~\cite{evlyukhin2011multipole, staude2013tailoring,anapole2019, terekhov2019broadband}, and it was revealed that the scattering properties of dielectric nanoparticles depend strongly on the geometric parameters and the value of the material refractive index.

According to Fig.~\ref{fig:properties}b, the refractive index of halide perovskites lies in the range $n=$2$\div$3, where compounds with narrower bandgaps ($\sim$1.5~eV, ABI$_3$) possess higher values around an exciton line, whereas lower values are associated with the compounds having the widest bandgaps ($\sim$3~eV, ABCl$_3$) satisfy qualitatively the Moss relation~\cite{moss1985relations}. PL of the halide perovskite nanoparticles follows the bandgap changing trends, and it can be varied in the entire visible frequency spectrum  (Fig.~\ref{fig:Blocks}d), which is important knowledge for Purcell factor calculations.

\subsection{Mie resonances}

Optical resonances in a dielectric spherical particle can be described analytically by the well-known Mie theory~\cite{mie1908beitrage}. The Mie theory gives the scattering efficiencies for all angles, as well as in both forward and backward directions relatively to the propagation direction of the incident plane wave:
\begin{equation}
  \label{eq:qscattot}
  C_{scat}^{total} = \frac{2}{q^2} \sum_{s = 1}^\infty (2s+1)(|a_s|^2 + |b_s|^2),
\end{equation}

\begin{equation}
  \label{eq:qscatfor}
  C_{scat}^{forward} = \frac{1}{q^2} |\sum_{s = 1}^\infty (2s+1)(a_s + b_s)|^2,
\end{equation}

\begin{equation}
  \label{eq:qscatback}
  C_{scat}^{backward} = \frac{1}{q^2} |\sum_{s = 1}^\infty (2s+1)(-1)^s(a_s - b_s)|^2,
\end{equation}
where, $q$=2$\pi R n_m /\lambda$, $R$ is the particle radius, $n_m$ is refractive index of host media, $\lambda$ is incident wavelength. The coefficients $a_s$ and $b_s$ correspond to electric- and magnetic scattering amplitudes for corresponding Mie modes (multipoles), which were actively studied during last decade for various high-refractive-index materials (such as Si and GaAs). 

The first resonance of a spherical dielectric nanoparticle is the magnetic dipole resonance ($b_1$), and it occurs when the wavelength of light inside a spherical particle with the refractive index becomes almost equal to the particle's diameter~\cite{Bohren1998small}. In turn, metallic spherical particles have negligibly small contribution of the magnetic modes. Another remarkable feature of dielectric particles is the ability to efficiently control scattering power pattern. As one can see from Eq.~(\ref{eq:qscatback}), backward scattering can be suppressed when magnetic and electric moments have comparable amplitudes and equal phases.

More advanced numerical methods are required to solve the Maxwell equations for cylindrical, cubical, and other shapes of nanoparticles~\cite{evlyukhin2011multipole}.
Despite the typical shapes of halide perovskite nanoparticles are non-spherical, but also cubic and cylindrical ones, one can similarly consider scattering on these nanoparticles as a series of multipoles. The multipole moments can be calculated through the electric current density $\mathbf{J} = -i \omega \varepsilon_{0} (\varepsilon_{\rm r} - 1) \mathbf{E}$ induced in the particle located in air by a normally incident wave, where $\epsilon_0$ is the vacuum permittivity, $\epsilon_r$ is the relative permittivity of the particle material. $\mathbf{E}$ is the total electric field inside the particle which can be obtained by carrying out 3D electromagnetic simulations. Since the dimensions of our particles are comparable with a wavelength of the incident light we used the following expressions for the spherical multipole expansion beyond the long-wavelength approximation~\cite{ALAEE201817}:

\begin{equation}
\label{eq:ED}
\textbf{p} = \frac{-1}{i \omega}  \int \textbf{J} j_0(kr) d\mathbf{r} + \frac{k^2}{2} \int \left[ 3 ( \mathbf{r} \cdot \mathbf{J} )\mathbf{r} - r^2 \mathbf{J} \right] \frac{j_2(kr)}{(kr)^2} d\mathbf{r},
\end{equation}

\begin{equation}
\label{eq:MD}
\textbf{m} = \frac{3}{2} \int [ \textbf{r} \times \textbf{J}] \frac{j_1(kr)}{kr} d\textbf{r},
\end{equation}

\begin{multline}
\label{eq:EQ}
\hat{Q} = \frac{-3}{i \omega} \int [3 (\textbf{r} \otimes \textbf{J} + \textbf{J} \otimes \textbf{r} ) - 2 (\textbf{r} \cdot \mathbf{J}) \hat{U} ] \frac{j_1(kr)}{kr} d \mathbf{r} \\
+ 2k^2 \int [5 \mathbf{r} \otimes \mathbf{r} (\textbf{r} \cdot \mathbf{J}) - (\mathbf{r} \otimes \mathbf{J} + \mathbf{J} \otimes \mathbf{r})r^2 - r^2(\mathbf{r} \cdot \mathbf{J})\hat{U}] \frac{j_3(kr)}{(kr)^3} d\mathbf{r},
\end{multline}

\begin{equation}
\label{eq:MQ}
\hat{M} = 15 \int [\mathbf{r} \otimes (\mathbf{r} \times \mathbf{J}) + (\mathbf{r} \times \mathbf{J}) \otimes \mathbf{r}] \frac{j_2(kr)}{(kr)^2}d\mathbf{r},
\end{equation}
where $\mathbf{p}$, $\mathbf{m}$, $\hat{Q}$ and $\hat{M}$ are electric dipole (ED), magnetic dipole (MD), electric quadrupole (EQ) and magnetic quadrupole (MQ) moments of the particle in the Cartesian representation, respectively. Here $j_{0,1,2,3}$ denotes spherical Bessel functions, $k$ is the wavenumber in air, $\hat{U}$ is the $3\times3$ unit tensor and $\mathbf{r}$ is the radius vector from the center of the particle.

Total scattering efficiency can be obtained by summing up each multipole contribution to the scattering as follows (here we consider contributions of several first multipoles up to magnetic quadrupole):
\begin{equation}
\label{eq:scatdecompos}
C_{scat} \simeq \frac{k^4}{6 \pi \epsilon_0^2 |\mathbf{E}_{inc}|^2 \sigma_{geom}} \left( \lvert \mathbf{p} \rvert ^2 + \frac{ \lvert \mathbf{m} \rvert ^2}{c}
+ \frac{\lvert k \hat{Q} \rvert ^2 + \left\lvert \frac{k \hat{M}}{c} \right\rvert ^2}{120} \right),
\end{equation}
where $\mathbf{E}_{inc}$ is the incident electric field and $\sigma_{geom}$ is the geometric cross section. 

In Fig.~\ref{fig:Mie}, scattering efficiencies for spheroids, cuboids, and cylinders made of different halide perovskite compositions are presented. The compositions correspond to three main formulas: MAPbI$_3$, MAPbBr$_3$, and MAPbCl$_3$. One can see that for the wavelengths larger than the edge of conduction band, Mie resonances are strong enough to provide high values of scattering efficiency (up to 10). At the wavelengths corresponding to high linear absorption, the scattering efficiency is much weaker except that for MAPbI$_3$, where the losses in range 500--800~nm (bottom of the conduction band) are not too high. Remarkably, that all considered shapes of perovskite nanoparticles demonstrate distinguishable Mie resonances.

\begin{figure}
\centering
\includegraphics[width=0.99\linewidth]{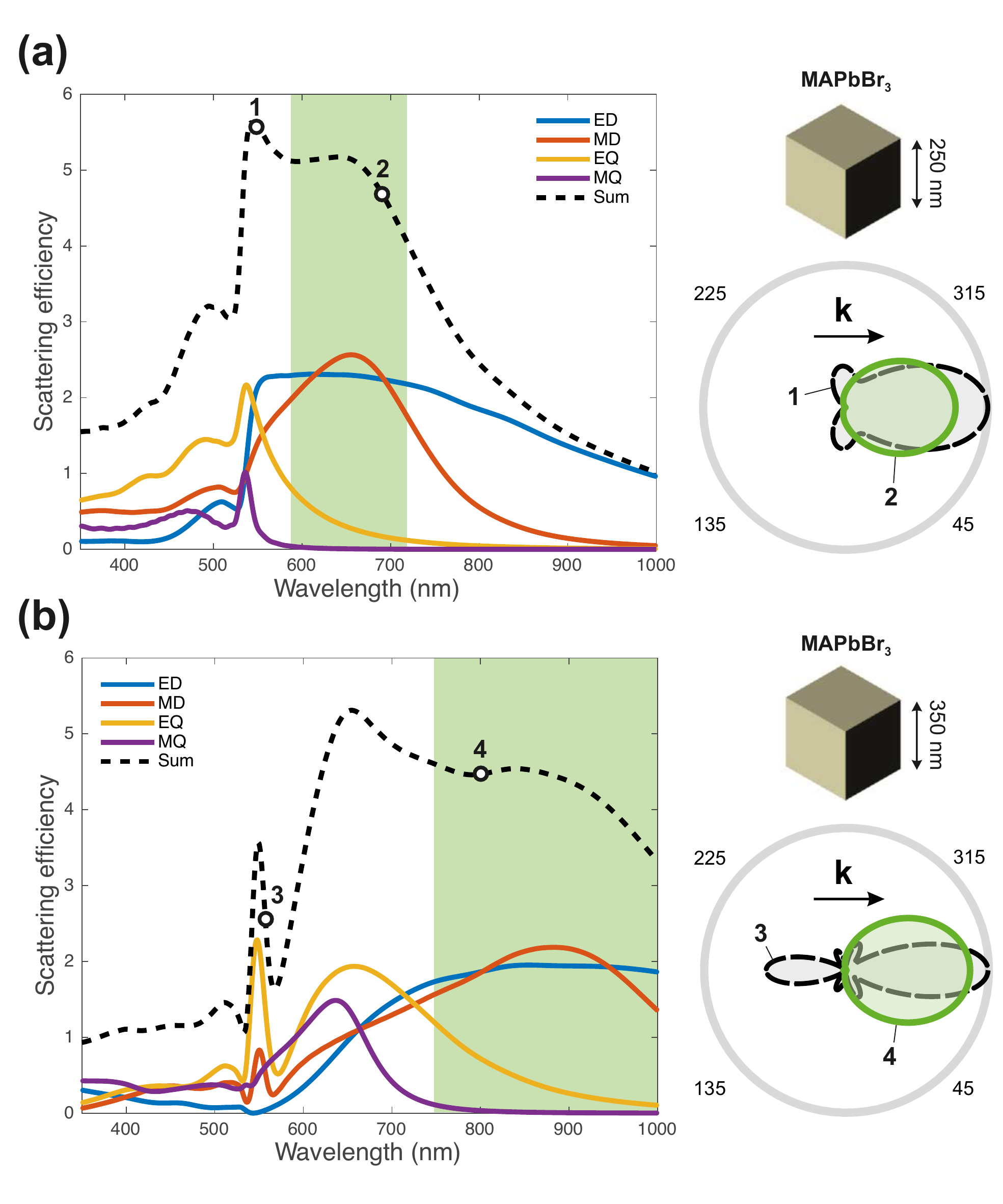}
\caption{\label{fig:Multipoles} Multipole decomposition of the scattering fields
of perovskite nanoparticles. Calculated total scattering and multipole decomposition from Eq.~(\ref{eq:ED} - \ref{eq:scatdecompos}) for MAPbBr$_3$ nanocubes with the size (a) $H = 250$~nm and (b) $H = 350$~nm. The green area corresponds to the Huygens' regime when ED $\simeq$ MD. Right: scattering patterns corresponding to the superposition of different multipoles: ED+EQ (point 1), ED+MD (points 2, 4) and EQ+MQ (point 3).}
\end{figure}

Next step is the analysis of scattering power pattern for resonant dielectric nanoparticles irradiated by a plane wave. Smart engineering of optical resonances within single nanoparticle by varying its geometrical parameters allows for interesting effects related to manipulation by scattering directivity. If electric and magnetic polarizabilities of a nanoparticle are equal each other in magnitude and phase (the first Kerker condition~\cite{kerker1983electromagnetic} or Huygens regime~\cite{decker2015high}), light scattering from this nanoparticle is suppressed in the backward direction. In this regard, it is crucial to provide multipole decomposition to understand at which spectral positions one can expect the regime of backward scattering suppression. In Fig.~\ref{fig:Multipoles}, the calculations show how different multipoles contribute to the scattering in a perovskite \textit{cubic} nanoparticle. As expected, the interference of electric and magnetic dipoles gives dominating forward scattering. Interestingly, that this regime is broadband for a specific sizes (e.g. for 350-nm cube, as shown in Fig.~\ref{fig:Multipoles}b), which is important for a number of applications related to anti-reflective coatings in real devices.

\subsection{Other resonances in dielectric meta-optics}

In addition to the well-known Mie resonances supported by individual meta-atoms, the recent advances in dielectric meta-optics are associated with other types of resonances which allow to shape and control both linear and nonlinear properties of nanoscale structures and metasurfaces.

The so-called {\it Fano resonance} is a fascinating phenomenon of wave physics observed in many problems of photonics, plasmonics, and metamaterials~\cite{fano_review}. A cluster of nanoparticles can support Fano resonances arising from interference between different localized modes and radiative electromagnetic waves. In nanophotonics, the Fano resonance is observed typically as a resonant suppression of the total scattering cross-section with enhanced absorption. The Fano resonance is attractive for applications because it allows to confine light more efficiently, because it is  characterized by a steeper dispersion than a conventional resonance, making it promising for nanoscale biochemical sensing, fast switching, and lasing.

A multimode analysis (see, e.g., the review~\cite{kruk} and references therein) reveals a principal difference between Fano resonances in dielectric and plasmonic nanoparticle clusters (also called oligomers). In comparison with oligomers made of metallic particles, all-dielectric nanoparticle structures are less sensitive to a separation between the particles, and the field is localized mainly inside the nanoparticles. Also, dielectric nanostructures exhibit low losses, which allows observing the Fano resonances for new geometries not supported by their plasmonic counterparts. These novel features lead to different coupling mechanisms between nanoparticles making all-dielectric nanoparticle structures more attractive for applications in active meta-optics. 

{\it Bound states in the continuum} have attracted a lot of attention in photonics recently, and they originate from a coupling between the leaky modes in dielectric structures such as photonic crystals, metasurfaces, and isolated resonators~\cite{BIC_review}. These resonances provide an alternative mean to achieve very large quality factors ($Q$-factors) for lasing~\cite{BIC_nature} and also allow to tune a photonic system into the regime of the so-called {\it supercavity mode}~\cite{rybin}. A true bound state in the continuum (BIC) is a mathematical object with an infinite value of the $Q$ factor and vanishing resonance width, and it can exist only in ideal loss-less infinite structures or for extreme values of parameters~\cite{hsu,alu}. In practice, BIC can be realized as a quasi-BIC mode, being directly associated with the supercavity mode~\cite{rybin}, when both the $Q$ factor and resonance width become finite. However, the localization of light inspired by the BIC physics makes it possible to realize high-$Q$ quasi-BIC modes in many optical structures such as cavities, photonic-crystal slabs, and coupled waveguides.

Importantly, there exists a direct link between quasi-BIC states and Fano resonances since these two phenomena are supported by the similar physics. More specifically, quasi-BIC resonance can be described explicitly by the classical Fano formula, and the observed peak positions and linewidths correspond exactly to the real and imaginary parts of the eigenmode frequencies. The Fano parameter becomes ill-defined at the BIC condition, which corresponds to a collapse of the Fano resonance. Importantly, every quasi-BIC mode can be linked with the Fano resonances, whereas the opposite is not always true: the Fano resonance may not converge to the BIC mode for any variation of the system parameters.

\subsection{Photoluminescence and lasing}

Optical resonances in nanoparticles is a powerful platform for the emission enhancement from nanostructures. Generally speaking, halide perovskites are considered as a family of materials with a high quantum yield (QY) of PL, being widely employed for LEDs and lasing applications, as described in Section~III above. However, the dependence of QY on photogenerated carriers concentration has a nontrivial behavior, which is strongly affected by the type of recombining particles: free carriers or excitons. Indeed, since binding energy of Wannier-Mott excitons in these materials is strongly dependent on the bandgap: $E_b$=20~meV for CsPbI$_3$, $E_b$=5$\div$15~meV for MAPbI$_3$, $E_b$=40~meV for CsPbBr$_3$, $E_b$=15$\div$40 meV for MAPbBr$_3$, $E_b$=60$\div$75~meV for CsPbCl$_3$, and $E_b$=41~meV for MAPbCl$_3$~\cite{jiang2019properties, makarov2019halide} free carriers and excitons can coexist in equilibrium under the appropriate conditions.

\begin{figure}[b!]
\centering
\includegraphics[width=0.9\linewidth]{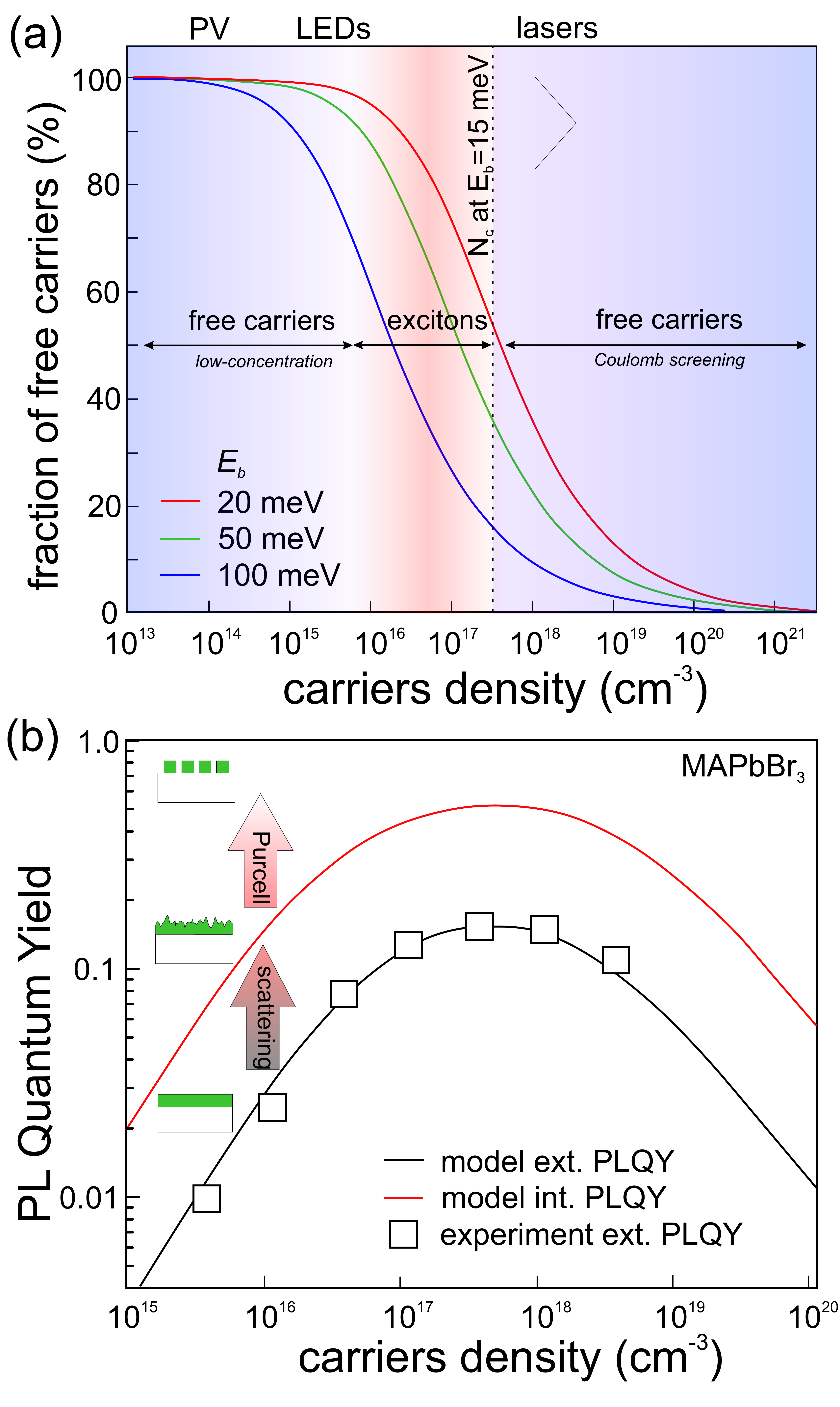}% Here is how to import EPS art
\caption{\label{fig:QY} (a) Calculated fraction of free carriers relatively to excitons as a function of injected carriers density according to the Saha–Langmuir equation for various values of exciton binding energies at room temperature. Carriers densities relevant for photovoltaic (PV), LEDs and lasing applications are highlighted. Adapted from Ref.~\cite{manser16}. (b) External PL quantum yield (PLQY) for MAPbBr$_3$ under pulsed laser excitation (squares) together with the modelled internal (red) and external (black) \textit{QY}. Insets illustrate schematically the origin of \textit{QY} enhancement and possible further improvement owing to Purcell effect in resonant perovskite nanoparticles. Adapted from Ref.~\cite{richter2016enhancing}.}
\end{figure}

A balance between free carrier and exciton populations at equilibrium can be determined by the Saha–Langmuir equation~\cite{saha1921physical, d2014excitons}. Fig.~\ref{fig:QY} shows the calculated fraction of free carriers relatively to excitons according to this equation with parameters close to 3D halide perovskite ones at room temperature taken from Ref.~\cite{manser16}. The calculated ratios for binding energies 20~meV (iodine-like), 50~meV (bromine-like), and 100~meV (chlorine-like) give general picture about the contribution of excitons for various compositions and regimes of perovskite excitation. 

However, the Saha–Langmuir relation does not include the contributions from many-body effects and can be applied at relatively low carriers densities only. Indeed, the Coulomb interaction between an electron and hole would be screened by the presence of other carriers, leading to the dissociation of excitons to free carriers. This transition (the so-called Mott transition) occurs at a critical excitation density ($N_{c}$) described by $N_{c}=k_BT/(11\pi E_ba_b^3)$, where $k_B$ is the Boltzmann constant, $T$ is temperature, $E_b$ is the exciton binding energy, and $a_b$ the Bohr radius. The dashed line in Figure~\ref{fig:QY}a represents the critical density ($N_{c}$ $\approx$ 4$\times$10$^{17}$cm$^{−3}$ ) of the Mott transition for $E_b$=15~meV and $a_b$=5~nm. The larger $E_b$ and smaller $a_b$ of the excitons, the higher $N_{c}$ for the excitons dissociation. In Fig.~\ref{fig:QY}a, this trend for the Mott transition toward higher $N_c$ is depicted by an arrow.

These estimations help to conclude that free carriers dominate both at low and high concentrations. The low-concentration case ($\approx$10$^{14}$$\div$10$^{16}$cm$^{-3}$) corresponds to the regimes of optoelectronic devices operation, while high-concentration case ($>$10$^{18}$cm$^{-3}$) corresponds to lasing at very strong pumping conditions.
In turn, excitons contribute considerably at intermediate carriers densities  ($\approx$10$^{16}$$\div$10$^{18}$cm$^{-3}$) for halide perovskites, depending on the binding energy $E_b$.
close to the thresholds of amplified spontaneous emission ($\approx$10$^{17}$$\div$10$^{18}$cm$^{-3}$). The coexistence of excitons and free-carriers in this regions raises debates on lasing mechanism in 3D halide perovskites~\cite{evans2018continuous, zhang2018strong, schlaus2019lasing, wei2019recent}.

Information about the dominating type of carriers is also crucial for the value of the PL QY because free-carriers and excitons have different relaxation behavior. The decay of photogenerated free carriers in perovskites is usually described with the following equation: 
\begin{equation}
\frac{dN_{fc}}{dt} = G - k_1 N_{fc} - k_2 N_{fc}^2 - k_3 N_{fc}^3,
\end{equation}
where $G$ is the photo-excited carriers generation rate. $k_1$ is the rate constant associated with trap-assisted recombination that relies on an individual carrier (electron or hole) being captured in a trap. The second relaxation term with $k_2$ corresponds to intrinsic electron–hole recombination, which depends on both electron and hole densities and, thus, on $N_{fc}^2$. $k_3$ is coefficient for non-radiative three-particle Auger recombination. As a result, the internal $QY$ can be written as: 
\begin{equation}
QY = \frac{k_2N_{fc}}{k_1 + k_2 N_{fc} + k_3 N_{fc}^2}
\end{equation}
More detailed analysis with taking into account exiton
contribution and defects see in Ref.~\cite{stranks2014recombination}.

Figure~\ref{fig:QY}b confirms that the internal QY of MAPbBr$_3$ depends almost linearly on photoexcited carriers density in the low-concentration regimes, saturating before the lasing regime owing to Auger recombination~\cite{richter2016enhancing}. Remarkably, the external QY from the perovskite film is much lower than the internal one, because emitted light can not escape the smooth film at high angles relatively to the normal to surface owing to strong reflection and re-absorption. Therefore, the enhancement of light scattering improving probability of emission outcoupling is one of the straightforward strategies for external QY improvement up to the level 30-90\% at maximum point~\cite{deschler2014high, richter2016enhancing, braly2018hybrid,cao2018perovskite}. However, the external QY still can be less than 10\%, when carriers concentration is relatively low. Therefore, smart nanophotonic designs further improving QY of luminescence are desirable for high-performance photonic and optoelectronic devices~\cite{lin2018perovskite}.

Beside the improving of the out-coupling, luminescence quantum yield can be enhanced because of acceleration spontaneous emission in a cavity $\tau_{sp, cav}$ as compared with that in homogeneous medium $\tau_{sp, 0}$ due to the Purcell effect~\cite{zambrana2015purcell}. The Purcell factor reaches then its maximal value $PF$ which in case of closed resonators has the form:
\begin{equation}
PF = \frac{\tau_{sp, cav}^{-1}}{\tau_{sp, 0}^{-1}} = \frac{6\pi c^3}{n^3\omega_m}\frac{Q}{V_{mode}},
\end{equation}
where \textit{n} is the refractive index, \textit{c} is speed of light in vacuum, $\omega_m$ is the angular frequency of the mode, and $V_{mode}$ is the mode volume. According to this formula, the increase of the $Q$-factor at fixed mode volume results in the acceleration of spontaneous emission rate, and, thus, the emitted power. Also, the increase of refractive index of the nanoparticle material results in better light localization and, thus, in \textit{Q}-factor enhancement and reduction of the mode volume. Thus, the Purcell effect is more pronounced for perovskites with higher refractive index, i.e. with narrower band gap (see Fig.~\ref{fig:Mie}b,f,j).

Taking into account spontaneous emission acceleration via the Purcell effect, the internal QY can be modified as following~\cite{tonkaev2019optical}:
\begin{equation}
QY = \frac{PF \cdot k_2 N_{fc}}{k_1 + PF\cdot k_2 N_{fc} + k_3 N_{fc}^2}
\end{equation}
According to the calculations in Fig.~\ref{fig:Mie} and Fig.~\ref{fig:QY}b, such a modification of spontaneous emission rate in resonant perovskite NPs should result in exceeding 10~\% level of external QY even for carriers density $<$10$^{16}$cm$^{-3}$, i.e. in the regime of optoelectronic devices operation.

\begin{figure}
\centering
\includegraphics[width=0.99\linewidth]{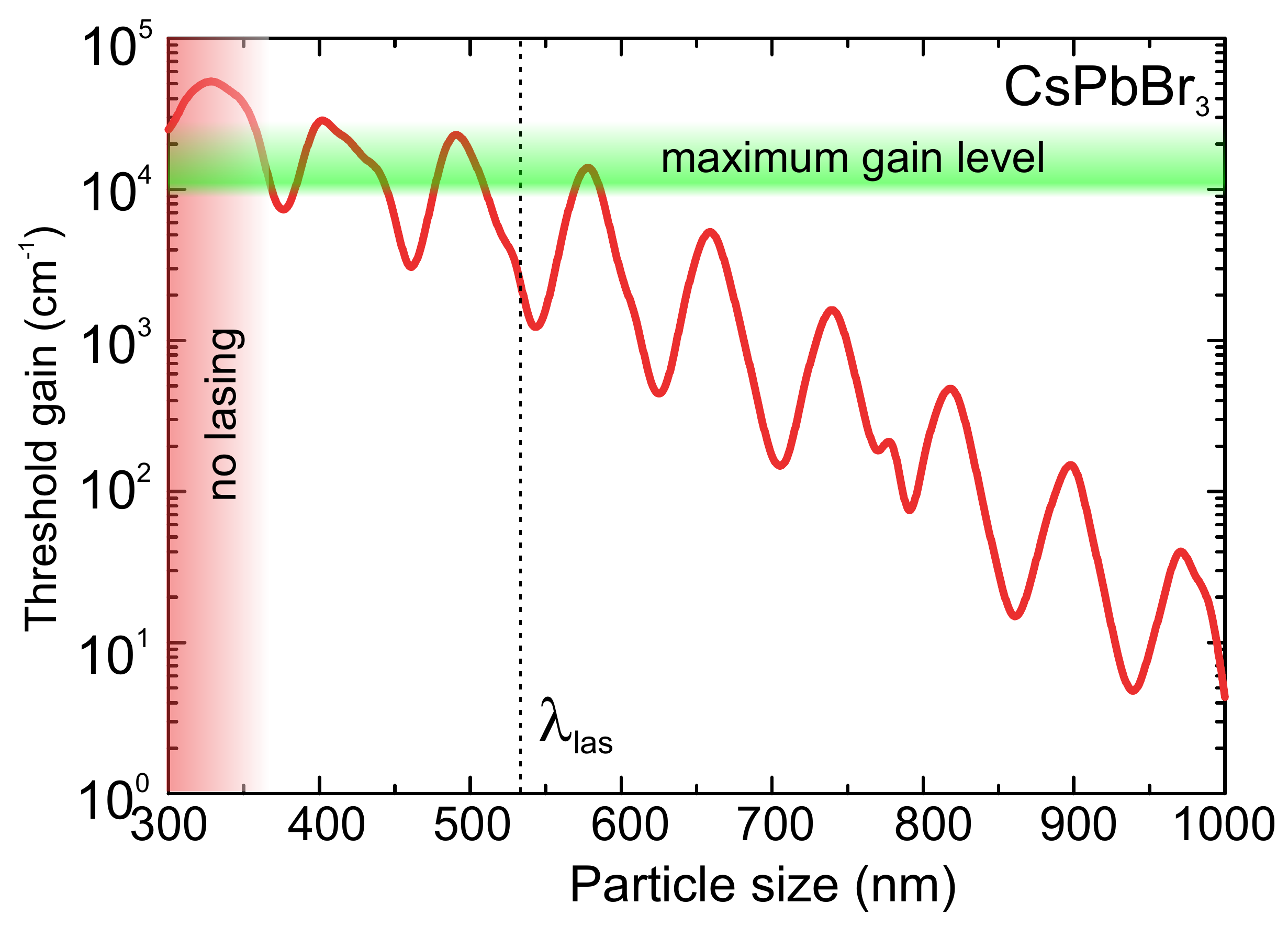}% Here is how to import EPS art
\caption{\label{fig:LasingTheory} Lasing of perovskite nanoparticles.  Dependence of the threshold gain coefficient on the size of CsPbBr$_3$ spherical nanoparticles in vacuum obtained from the linear theory. Green stripe corresponds to the maximum levels of gain in CsPbBr$_3$. Red region corresponds to nanoparticles with small $Q$-factors not supporting lasing. Adapted from Ref.~\cite{tiguntseva2019single}.}
\end{figure}

Higher modes in perovskites nanocavities can possess high enough \textit{Q}-factors to achieve \textit{lasing}. Generally, \textit{Q}-factor in a dielectric subwavelength cavity exponentially depends on the mode number \textit{m}:~\cite{shainline2009subwavelength}
\begin{equation}
Q(m) \sim e^{\beta m},
\end{equation}
where $\beta$ is the constant determined by the cavity geometry, optical properties of the cavity and host material. More advanced designs were recently proposed for high-refractive-index nanocylinders~\cite{rybin2017high}, where high-\textit{Q} modes were highly symmetric and badly out-coupled into freely propagating light in surrounding dielectric medium. This type of high-Q states is frequently called bound states in continuum (BIC). 

Eventually, optimization of \textit{Q}-factor is crucial for compactization of a microcavity and minimization of its threshold modal gain ($g_{\it th}$) for lasing: $g_{\it th}=2 \pi n/\lambda Q_{\it th}$. Fig.~\ref{fig:LasingTheory} shows calculated dependencies of thresholds for spherical CsPbBr$_3$ nanoparticles on their size (diameter for sphere and edge length for cube). Remarkably, high gain in CsPbBr$_3$ perovskites (up to 3$\times$10$^4$cm$^{-1}$ depending on temperature) still allows for lasing in subwavelength nanoparticles ($<$540~nm) regardless their shape, because low-order Mie resonances are not so sensitive to the shape. Larger particles demonstrate exponential general trend for reduction of the lasing threshold [see also Eq.~(10)]. However, spherical particles are much more preferable when the size is larger than $\sim$800~nm, when whispering gallery modes represent increase of incidence angle of lasing mode to the spherical boundary of particle, whereas this angle is fixed for the cubic particles. Taking into account relatively low thermal conductivity of halide perovskites ($<$10~W/m$\cdot$K)~\cite{sutherland2016perovskite}, halide perovskite micro-particles ($>$1~$\mu$m) of cubic shape are much worse as compared to more compact sub-micron spherical particles with similar $Q$-factors. Nevertheless, final decision on the design selection is based on technological and experimental aspects, which we discuss in the next Section.

\section{Fabrication methods}

\subsection{Resonant nanoparticles} 

In order to study novel physics of active meta-optics based on the Mie-type modes in halide perovskites, various building-block designs were fabricated. Namely, in the most of experiments, the perovskite nanoparticles with sizes smaller than 1~$\mu$m acquired spherical, cubic, and cylindrical shapes.

Chemical methods allow to create randomly arranged perovskite nanoparticles with cubic shapes mostly. For example, the ligand assisted reprecipitation method~\cite{huang2015control} or antisolvent-assisted techniques in aprotic non-polar liquids give a cubic shape of nanoparticles with a good crystalline structure and sizes ranging from a few to hundred nanometers~\cite{liu2018robust,berestennikov2019beyond}. 
Synthesis of nanoparticles with non-cubic (polygon) shapes were achieved by a polymer-assisted precipitation on a substrate after spin-coating and annealing of precursors~\cite{tiguntseva2018tunable}.

Another technique is a chemical vapour deposition method allowing for fabrication of microcubes~\cite{zhou2018single} and microspheres~\cite{tang2017single} of halide perovskites randomly arranged on a substrate. This method is specifically prospective for creation of high-quality spherical particles~\cite{tang2017single}, which is difficult to achieve by the other methods.

Creation of perovskite nanoparticles with a quasi-spherical shape and much higher precision of spatial positioning was demonstrated by means of laser transfer technique~\cite{tiguntseva2018light, tiguntseva2018tunable}. It represents a physical method (laser ablation) to obtain single perovskite nanoparticles flying from spin-coated perovskite films to an arbitrary substrate. Also, direct laser ablation of a thin perovskite film allows for cutting microstructures with shapes (e.g. cylinders), which are not possible to produce by any chemical or chemical vapor deposition methods~\cite{zhizhchenko2019single}. Remarkably, that the high defect tolerance of halide perovskites preserves high quantum yield for luminescence from the fabricated nanoparticles or microdisks even after the intense laser-matter interaction process.

Finally, perovskite microparticles of a cylindrical shape can be fabricated by electron- or ion-beam lithography with the highest spatial precision~\cite{alias2015enhanced,zhang2017highly}. For example, microdisks with various shapes were created and even integrated with nanoscale waveguide~\cite{cegielski2018monolithically}.

\subsection{Metasurfaces} 

Metasurfaces are periodical nanostructures consisting of the nanoparticles of any shapes and arranged with the periods less than incident or emitted light wavelengths. Beside the special requirements for nanoparticles shape, the periodicity in metasurfaces is another critical parameter. Therefore, it is very challenging to create high-quality metasurfaces by means of any chemical methods, including various template-assisted techniques~\cite{duan2018chip}. The highest precision was achieved by employing ion-beam~\cite{gholipour17} and electron-beam~\cite{chen2016photonic, gao2018lead, wang2018lead} lithography. However, these lithography-based approaches suffer from relatively low productivity. In this regard, nanoimprint lithography demonstrated high potential for mass-production of halide perovskite photonic crystals and metasurfaces~\cite{wang2016nanoimprinted, makarov2017multifold, wang2017nanoimprinted,tiguntseva2017resonant}.   

\section{Experimental demonstrations}

\subsection{Luminescence enhancement in resonant nanoparticles} 

\begin{figure}
\centering
\includegraphics[width=1.0\linewidth]{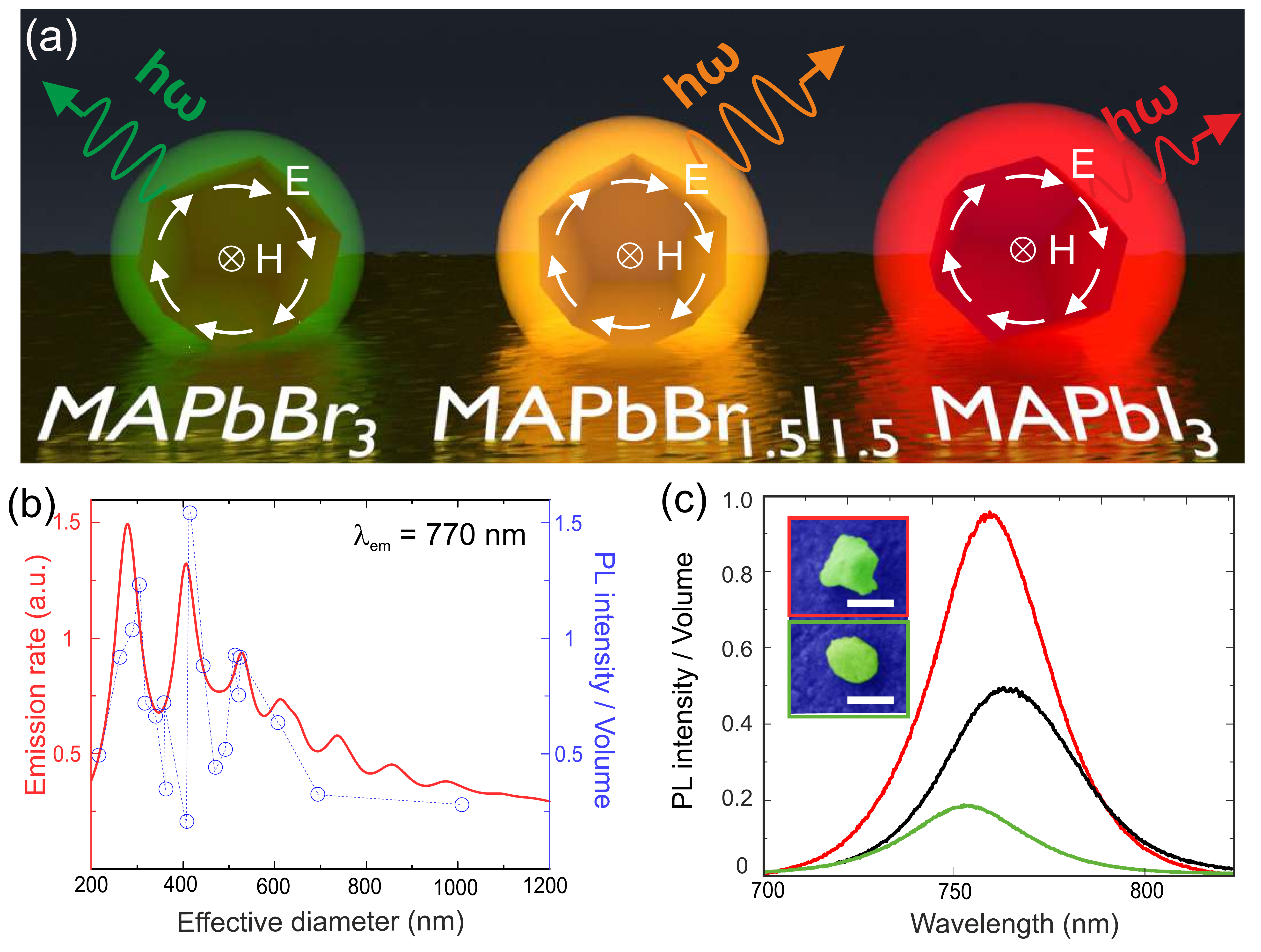}% Here is how to import EPS art
\caption{\label{fig:PL} Resonant enhancement of photoluminescence in halide perovskite nanoparticles. (a) Schematic illustration of halide perovskite nanoantennas. (b) Experimental dependence of photoluminescence signal on the size of MAPbI$_3$ nanoparticles and corresponding theoretical calculations for the emission rate. (c) Experimental spectra of photoluminescence from MAPbI$_3$ nanoparticles of different sizes and thin MAPbI$_3$ film. Adopted from the paper~\cite{tiguntseva2018light}.}
\end{figure}

In the work~\cite{tiguntseva2018light}, pronounced Mie resonances were observed experimentally in individual nanoparticles smaller than 500~nm made of different types of halide perovskite MAPb(Br$_x$I$_{1-x}$)$_3$, as shown schematically in Fig.\ref{fig:PL}a. As we discussed in previous sections, the main advantages of Mie resonances are local field enhancement and far-field control for incident light, as well as acceleration of radiative recombination at the emission wavelength of material. The latter effect results in the enhancement of luminescence efficiency around the Mie resonances, which was also observed experimentally for halide perovskite nanoparticles. In Fig.\ref{fig:PL}b, the perovskite nanoparticles made of MAPbI$_3$ exhibited clear dependence on their size, which can be well fitted qualitatively by analytical calculations based on Eq.(11), or more rigorously basing on the Chew's theory.~\cite{chew1988radiation} The PL from the nanoparticle with the optimized size is up to 5 times stronger as compared with a thin film (see Fig.\ref{fig:PL}c). However, optical losses in perovskites at the exciton peak and increasing mode volume with a growth of particle's size reduces an averaged emitted power, being maximum for the magnetic quadrupole resonance.~\cite{tiguntseva2018light}

\begin{figure}
\centering
\includegraphics[width=1.0\linewidth]{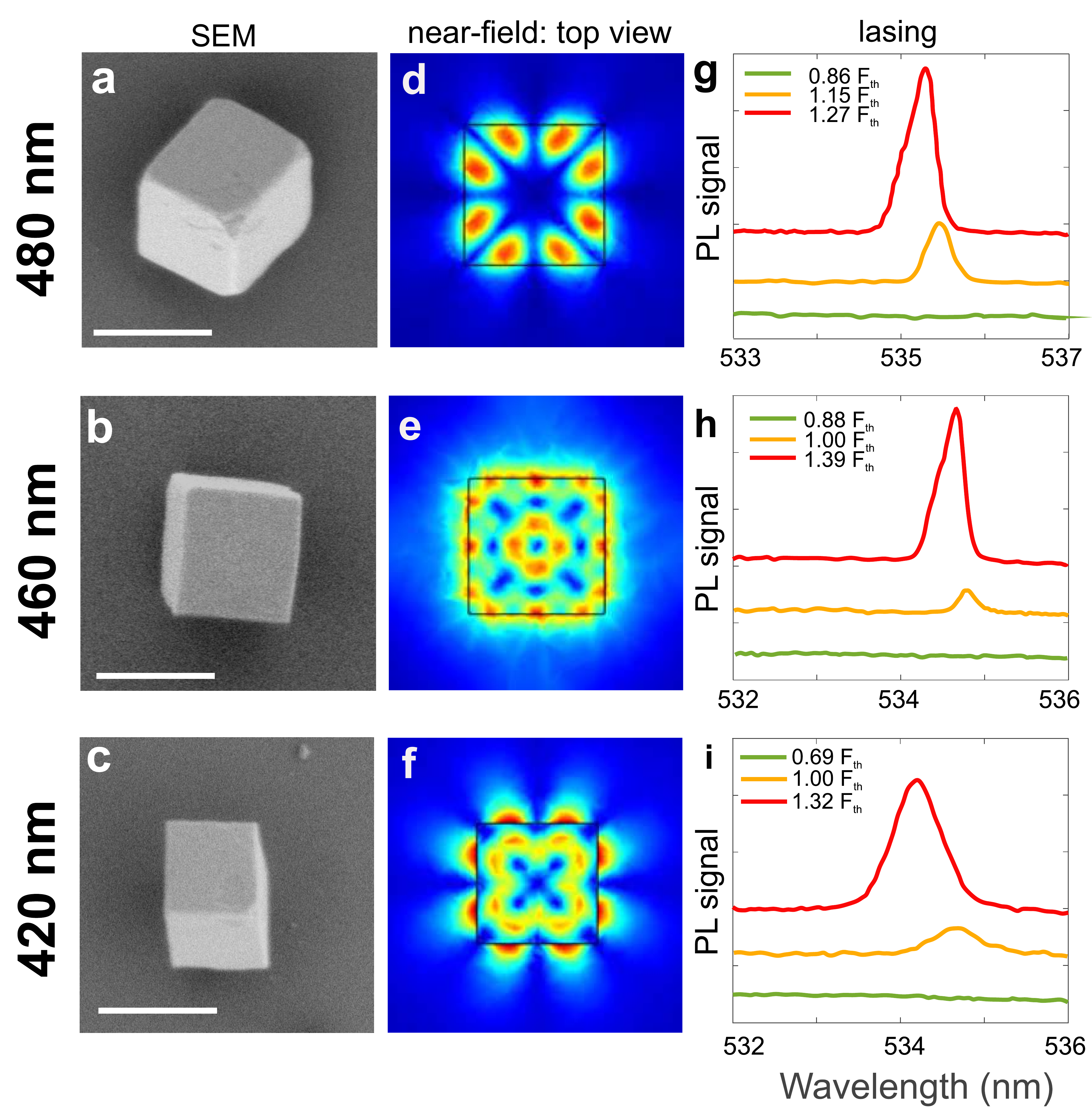}% Here is how to import EPS art
\caption{\label{fig:Lasing} Lasing from perovskite nanoparticles. (a-c) Scanning electron microscopy (SEM) images of CsPbBr$_3$ perovskite nanocubes on a sapphire substrate. Corresponding sizes are shown next to each image. Scale bars are 500~nm. (d-f) Calculated electric field distribution of the lasing mode (around 535~nm) in the nanocubes. (g-i) Experimentally measured dependencies of the emission intensity from the nanocubes at room temperature. Adapted from Ref.~\cite{tiguntseva2019single}.}
\end{figure}

\begin{figure*}
\centering
\includegraphics[width=0.9\linewidth]{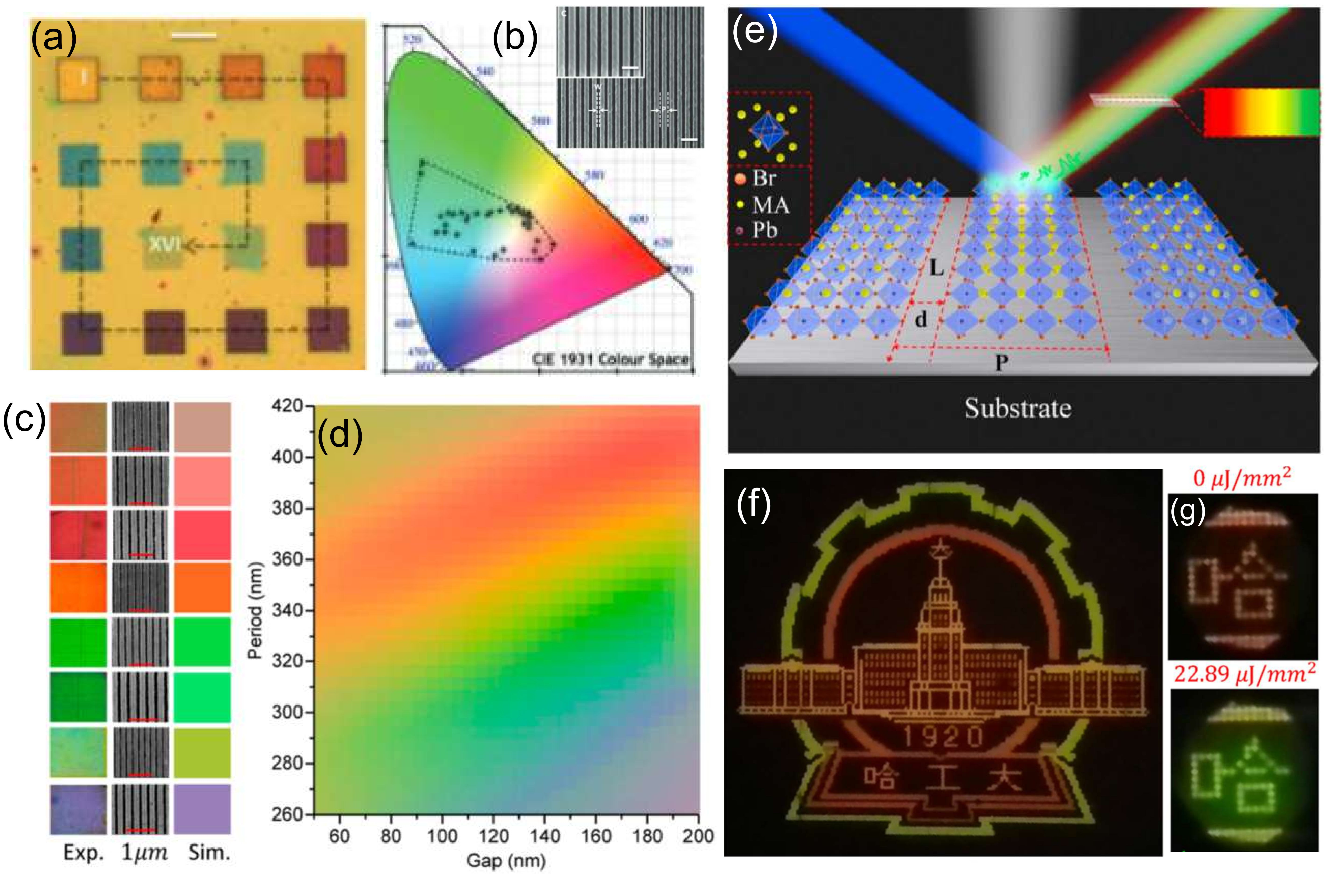}% Here is how to import EPS art
\caption{\label{fig:color} Coloring with perovskite metasurfaces and nanostructures. (a) Optical images of MAPbI$_3$ metasurfaces with different periods and (b) corresponding CE 1931 color space for the meta-pixels with an electron image. (c) Optical and electronic images of perovskite metasurfaces, as well as corresponding simulated colors. (d) Reflection colors for perovskite metasurfaces. (e)  Perovskite metasurfaces with actively tunable colors by means of luminescence excitation and its experimental demonstrations (f) without and  (g) with photoluminescence contribution.~\cite{gao2018lead}
}
\end{figure*}

\subsection{Excitonic effects in resonant nanoparticles} 

Beside the PL enhancement effect, it is expected that coherent coupling of Mie modes of the particle to the excitons~\cite{platts2009whispering} of perovskite will result in a pronounced Fano resonance at room temperature, because the halide perovskites possess pronounced exciton state with binding energies more than 20~meV~\cite{tanaka2003comparative}. As show in work~\cite{tiguntseva2018tunable}, the calculated scattering cross section spectra of spherical  MAPbI$_3$ nanoparticles versus particle radius reveal an asymmetric behavior inherent for Fano-like dip close to the exciton resonance of MAPbBr$_3$ perovskite at 539~nm. This knowledge is crucial for the analysis of transmittance spectra for the films made of perovskite nanoparticles, where the dip around exciton spectral line has to be attributed to the Fano resonance rather than reduced absorption in the material.

Another interesting exciton-driven effect is related to a nonlocal character of dielectric permittivity of halide perovskites. The mechanism of this effect is associated with the fact that excitons in nanoparticles have an additional kinetic energy that is proportional to $k^2$, where $k$ is the wavenumber. Therefore, they possess higher energy than in the case of static excitons. In more details, spatial dispersion inside the sphere results in the emergence of the additional, longitudinal electric wave, $\mathbf{E}_l$.~\cite{Agranovich1965c} This wave coexists with the conventional, transverse waves and can also be expanded in spherical harmonics with a single set of expansion coefficients. As shown in the work~\cite{berestennikov2019beyond}, this phenomenon is important at the intermediate sizes of nanoparticles, where the nonlocal response of the exciton affects the spectral properties of Mie modes. As a result, the some blue-shift of the PL, scattering, and absorption cross-section peaks can be observed for halide perovskites with sizes in the range between 10 and 100~nm, i.e. where any quantum confinement effects are negligible, while the contribution Mie modes is considerable.

\subsection{Lasing from nanoparticles} 

According to the theoretical predictions (see Fig.~\ref{fig:LasingTheory}), slightly larger perovskite nanoparticles are prospective for lasing applications. Indeed, owing to the high QY~\cite{deschler2014high} and ability to support confined optical modes~\cite{makarov2019halide}, halide perovskite based nanoparticles are also very prospective for creation of micro- and nanolasers operating at room temperature~\cite{wei2019recent}. The first successful demonstration of this concept with sub-micron particles was carried out with perfectly spherical CsPbBr$_3$ particles with a diameter of about 780~nm~\cite{tang2017single}. Reduction of size resulted to significant increase of lasing threshold from 10$^{-1}$~$\mu$J/cm$^2$ level~\cite{tang2017single} up to 10$^2$~$\mu$J/cm$^2$ for sub-wavelength cubic nanoparticles~\cite{tiguntseva2019single}. As shown in Fig.~\ref{fig:Lasing}, the  CsPbBr$_3$  subwavelength nanocubes of a high quality can support an efficient lasing at the fourth-order Mie resonance (electric hexadecapole) even at room temperatures, being as small as 0.5~$\lambda^3$ and the smallest non-plasmonic nanolaser~\cite{tiguntseva2019single}.  Further size decrease results in exceeding the damage threshold of the material. 

\subsection{Structural and active coloring} 

The ordered resonant perovskite nanoparticles or resonant nanostructures (so-called `meta-atoms’) can form a metasurface. The first application of halide perovskite metasurfaces was structural coloration~\cite{gholipour17}, when different wavelengths of the incident light are selectively reflected and absorbed by the metasurface owing to its resonant response (Fig.~\ref{fig:color}a-b). Indeed, as shown in Fig.~\ref{fig:Mie}, perovskite nanoparticles can demonstrate enhanced scattering at various wavelengths because of Mie resonances. Some particular realizations of perovskite periodical nanostructures and corresponding structural colors are presented in Fig.~\ref{fig:color}c-d. Similar effect was demonstrated previously with other semiconductors materials (e.g., silicon~\cite{proust2016all} and germanium~\cite{zhu2017resonant}).

Additional color mixing becomes possible when incident light not only passively propagates through the metasurface, but also makes its active, i.e. induces photolumineacence or even lasing at a certain intensity and wavelength. Figure~\ref{fig:color}e represents an experimental illustration of the active metasurface based on MAPbBr$_3$ perovskite nanostructure, where different colors are obtained under different intensity of incident light with wavelength 400~nm~\cite{gao2018lead}. This paves the way for display pixels with tunable colours originated from resonantly enhanced luminescence and selectively modified reflection properties of halide perovskite metasurfaces.

\begin{figure*}
\centering
\includegraphics[width=1.0\linewidth]{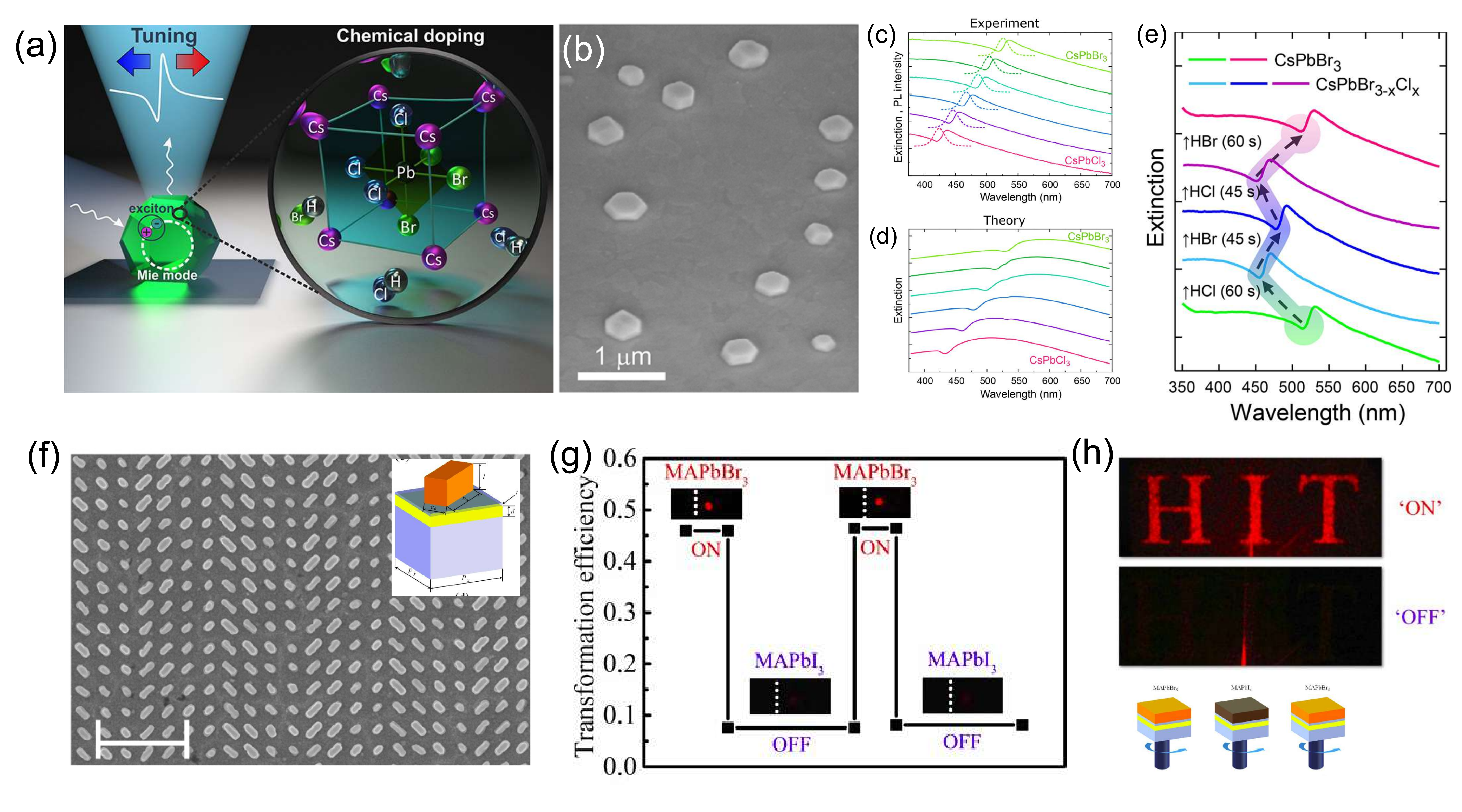}
\caption{\label{fig:tuning} Dynamically reconfigurable meta-atoms and metasurfaces. (a) Concept of nanoantenna tunability by an anion exchange in the vapor phase. (b) Electron image of CsPbBr$_3$ resonance nanoparticles that demonstrates (c) a gradual shift of the Fano resonance after the anion exchange process (d) confirmed by theoretical modeling~\cite{tiguntseva2018tunable}. (e) Experimental demonstration of the reversible multicycle tuning of the Fano resonance~\cite{tiguntseva2018tunable}. (f) Electron image of MAPbBr$_3$ metasurface switched to MAPbI$_3$ phase by an anion exchange, as shown in (g). (h) Holographic images before and after the chemical switching of perovskite metasurfaces shown schematically in the lower row of images.~\cite{Shumin2019dynamic}}
\end{figure*}

The remarkable feature which makes halide perovskites unique among the other semiconductors is that one can tune their optical properties \textit{in-situ} by the anion exchange~\cite{pellet2015transforming, solis2015post}, which is much more controllable than photo-induced segregation in perovskites~\cite{hoke2015reversible}. The anions (I$^-$, Br$^-$, and Cl$^-$) exchange in arbitrary perovskite objects (thin films, nanoparticles, nanowires, etc.) can occur in gas phase when they are surrounded by an acid molecules (HI, HBr, or HCl). This approach was applied to resonant CsPbBr$_3$ nanoparticles as well and resulted in a broadband ($\sim$1~eV) exciton spectral position tuning by applying HCl acid in gas phase by several minutes~\cite{tiguntseva2018tunable}. As shown in Fig.~\ref{fig:tuning}a-c, white light transmission spectrum through the randomly distributed resonant CsPbBr$_3$ perovskite nanoparticles contains the feature of Fano resonance caused by coupling the exciton with Mie modes in the nanoparticles, and it is gradually modified at different doses of HCl gas. Moreover, Fig.~\ref{fig:tuning}e proves that the nanoparticles spectrum can be reversibly modified upon exposure in vapours of HCl and HBr acids.

The effect of perovskite band-gap engineering was applied for strong spectral tuning of the optical properties of MAPbBr$_3$ metasurface (see example in Fig.~\ref{fig:tuning}f) by means of \textit{in situ} material reversible conversion to MAPbI$_3$ in a chemical vapor deposition (CVD) tube~\cite{Shumin2019dynamic}. This approach allowed for switching-off reflection signal reversibly from the perovskite metasurface (Fig.~\ref{fig:tuning}g). A switchable perovskite meta-hologram was also demonstrated (Fig.~\ref{fig:tuning}h), when the material of the design converting incident light into the word ``HIT'' was modified in the CVD tube.

\subsection{Nonlinear metasurfaces}

Nonlinear meta-optics based on Mie-resonant nanoparticles made of Si, GaP, Ge, or GaAs materials is a very rapidly developing area of modern optics~\cite{shadrivov2015nonlinear}. In turn, the use of halide perovskite nanoparticles and metasurfaces is also very attractive for enhancing nonlinear optical response of metadevices, as was discussed above.

\begin{figure*}
\centering
\includegraphics[width=1.0\linewidth]{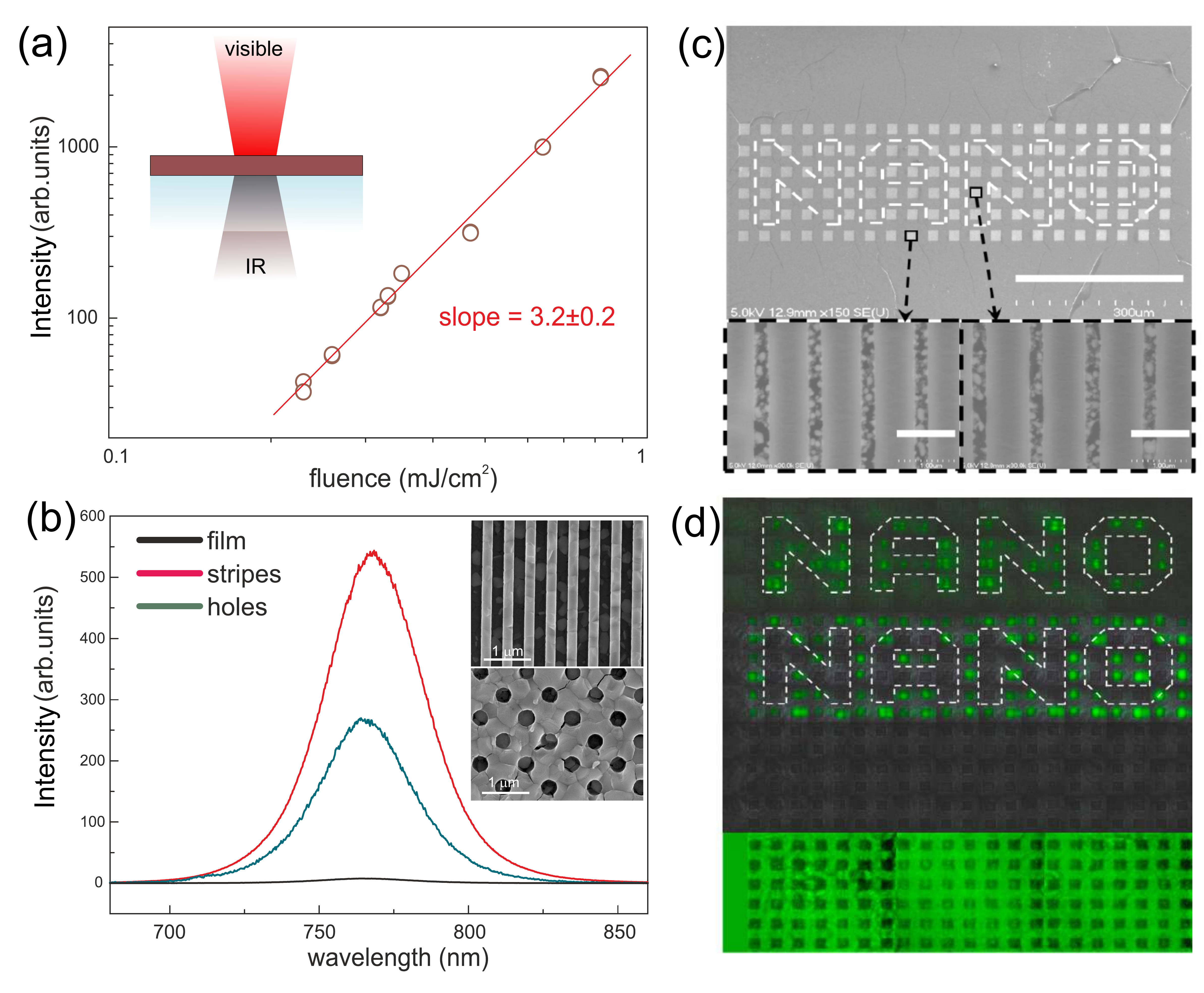}% Here is how to import EPS art
\caption{\label{fig:nonlinear} Nonlinear up-conversion from halide perovskite metasurfaces. (a) Experimentally measured dependence of nonlinear photoluminescence on incident fluence of 150-fs infrared laser pulses. (b) Experimental spectra of nonlinearly pumped photoluminescence from different triple-cation mixed-halide perovskite nanostructures with corresponding electron images.
(c) A top view electron images of MAPbBr$_3$ metasurface for nonlinear imaging, scale bar is 300~$\mu$m. The insets are the images of the encoded information and the background, scale bar is 1~$\mu$m. (d) Nonlinear photoluminescence images under different pumping wavelengths 1500~nm (upper), 1400~nm and 1350~nm of femtosecond laser, and linear photoluminescence image pumped by a laser at wavelength 400~nm (lower).~\cite{fan2019resonance} }
\end{figure*}

An increase up to two orders of magnitude in multiphoton absorption induced PL was demonstrated with perovskite metasurfaces pumped by near-infrared fs laser pulses~\cite{makarov17}. The enhancement factor depends on the type of nanostructure (see Fig.~\ref{fig:nonlinear}b), and a strong polarization dependence was demonstrated owing to excitation of various modes. Recently, such a strong enhancement was employed~\cite{fan2019resonance} to distinguish meta-pixels with slightly different designs or excitation wavelengths, as shown in Fig.~\ref{fig:nonlinear}c-d. This finding opens a new degree of freedom for structural coloring and encoding of information. Nonlinear metasurfaces made of plasmonic structures coupled to a perovskite layer were also proposed recently for ultrafast modulation of terahertz signals~\cite{chanana2018ultrafast, manjappa2017hybrid}, demonstrating picosecond-scale material`s response.

\section{Summary and outlook}

\begin{figure}
\centering
\includegraphics[width=1.0\linewidth]{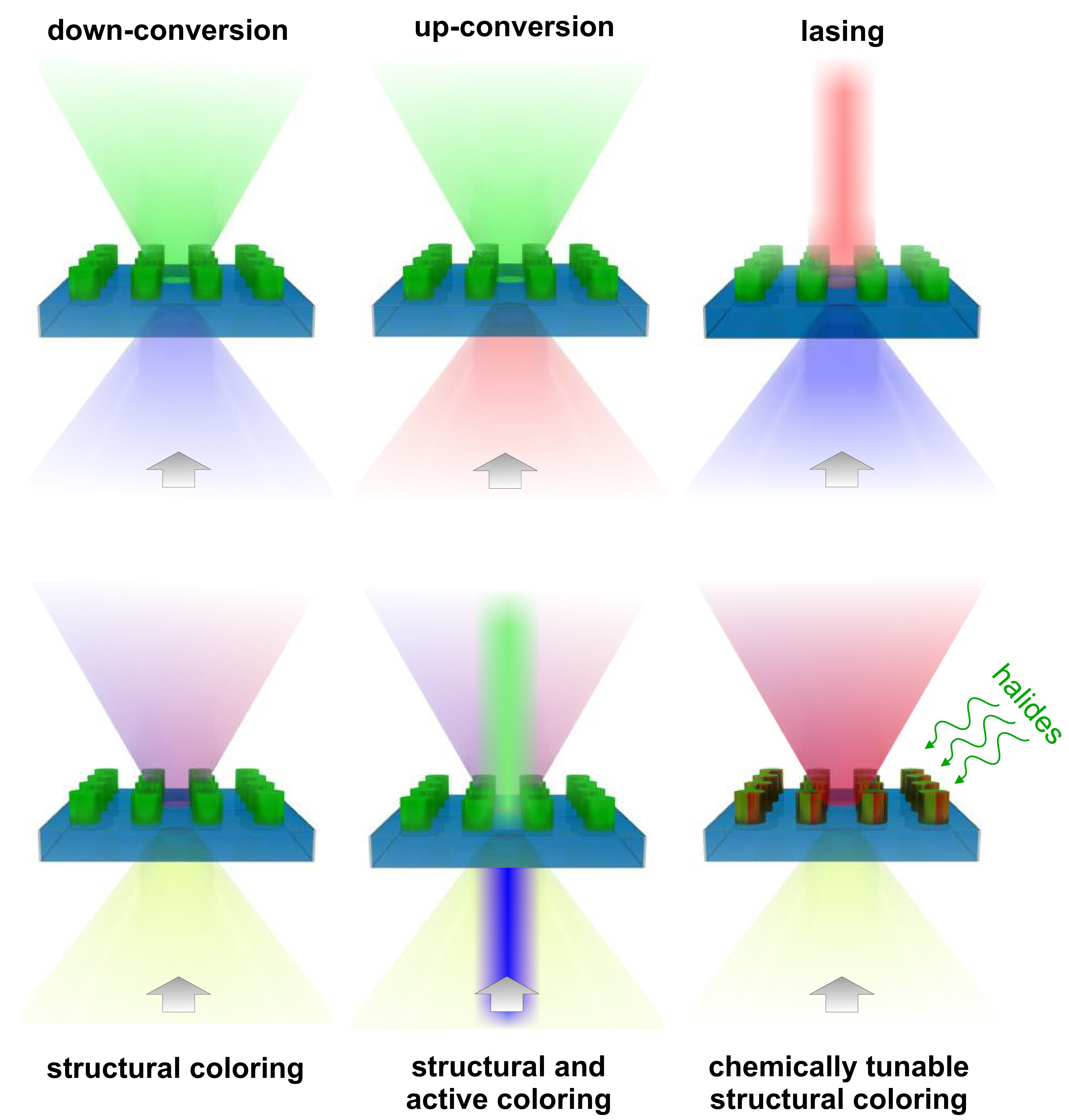}% Here is how to import EPS art
\caption{\label{fig:app} Schematic illustration of several different applications of perovskite meta-optics for nonlinear nanophotonics, nanolasers, resonance-based coloring, and chemically controlled tunability.
}
\end{figure}

We have presented a brief overview of the recent advances in the emerging field of active meta-optics and dielectric nanophotonics 
based on halide perovskites. In particular, we have demonstrated that nanoparticles made of halide perovskites and having various shapes can support pronounced Mie resonances in visible and near-infrared spectral range. Such optically-resonant perovskite nanoparticles can provide a subwavelength confinement of the local electromagnetic fields, and they can boost many optical effects originating from strong light-matter interaction and optical magnetism, thus offering novel opportunities for the subwavelength control of light in active and light-emitting nanostructures and metasurfaces. Figure~\ref{fig:app} summarizes several recently developed applications of halide-perovskite active and light-emitting metasurfaces. These examples range from a standard wave-front control and spectrally-resolved wave selection to novel functionalities such as highly-efficient luminescence, compact lasing, and reversible chemistry-induced tunability.

For the future developments, we emphasize that a combination of halide perovskites with nanostructuring suggest that the performance of many perovskite-based optical devices can be improved substantially, and perovskite devices can be empowered by optical resonances and nanostructuring. Below we outline several research directions where we oversee some further progress in the next coming years.  

{\it Ultimately small nanolasers}. According to Fig.~\ref{fig:LasingTheory}, one can expect room-temperature lasing at the sizes considerably smaller than the wavelength of emitted photons (down to 0.6$\lambda$) owing to a high gain of halide perovskites. Further decreasing the nanolaser size becomes possible by employing various strategies. The first strategy is a decrease of nanolaser temperature in order to simply increase the population of excited carriers at the bottom of the conduction band as usually happens in semiconductors~\cite{coleman2011semiconductor}. The second strategy is related to the optimization of perovskite material properties. For example, cooling halide perovskite can lead to phase transitions and phase co-existence at some temperature intervals (e.g. at $T<$160~K, for MAPbI$_3$). In such a case, a gain originates from tetragonal phase inclusions that are photogenerated by the pump within the bulk orthorhombic host matrix with slightly wider band gap and lower losses in the range of lasing from the tetragonal phase inclusions~\cite{jia2017continuous}. Similar multiphase optimization  might be achieved  potentially even at room temperatures by anion-mixing in perovskites, where a segregation process results in a separation of different mono-halide perovskite phases~\cite{slotcavage2016light} and, thus, the formation of inclusions with smaller bandgaps inside a host medium with a wider bandgap.

{\it Enhanced electroluminescence.}  Resulting from the previous demonstration of the enhanced PL from perovksite meta-atoms~\cite{tiguntseva2018light}, metasurfaces~\cite{gholipour17, makarov2017efficient, gao2018lead, tiguntseva2019enhanced, Shumin2019dynamic, fan2019resonance}, and also electroluminescence from random nanoparticles~\cite{cao2018perovskite} and photonic crystals~\cite{mao2017novel}, we anticipate the demonstration of electrically-driven metasurface-based perovskite LEDs. This would result in a dynamical control of LED emission and may offer new opportunities for efficient and smart displays. Remarkably, current state-of-the-art perovskite-based LEDs can be semi-transparent~\cite{lee2019flexibility}, which allows for an additional control of not only emitted light, but also transmitted light expanding the device functionality. Finally, the optimized perovskite metasurfaces supporting high-\textit{Q} modes can be utilized for creating electrically pumped perovskite lasers, which has not been demonstrated so far.

{\it Holographic luminescent or lasing images.} Recent progress in all-dielectric light-emitting nanoantennas and metasurfaces~\cite{vaskin2018directional, vaskin2019manipulation} offer new opportunities for control of the emission wavelength, brightness,  wavefront shape, and direction of the emitted light. This would be beneficial for 3D imaging or holography~\cite{yaracs2010state, wang2016grayscale}, where different encoded images are projected at different viewing angles. All-dielectric perovskite-based metasurfaces supporting lasing could also improve quality of imaging in holographic displays owing to high coherence and improved directivity of the emitted light.

{\it Perovskite photovoltaics.} Highly efficient perovskite solar cells with structural colors~\cite{eperon2013neutral, zhang2015highly} presents another important step for building integrated photovoltaic devices and smart glass. On the other hand, randomly distributed resonant plasmonic~\cite{zhang2013enhancement} and dielectric~\cite{furasova2018resonant} nanoparticles have already been utilized for boosting photovoltaic characteristics. Next logical step in this direction would be to combine structural colors in transmittance/reflectance, enhanced absorption, and even controlled luminescence  directivity to create multifunctional light-emitting solar cells~\cite{gets2019light} with an outstanding performance.

{\it Enhanced sensing}. Recent designs based on high-\textit{Q} all-dielectric  metasurfaces~\cite{tittl2018imaging, yesilkoy2019ultrasensitive} demonstrate outstanding functionalities of nanostructured devices for biosensing applications. In turn, halide perovskite metasurfaces with similar designs would allow creating  highly sensitive liquid~\cite{zhang2015highly}, gas~\cite{tiguntseva2018tunable}, and deformation~\cite{yang2019controllable} sensors. This would allow to acquire spatially resolved spectra from millions of image pixels and use smart data-processing tools to extract high-throughput digital sensing information at the unprecedented level of sensitivity
with the spectral data retrieval from a single image without using spectrometers.

\textit{Optical cooling}. The effect of temperature decrease upon material illumination by photons with the energies lower than that of the excitonic state (upconversion) and absorption of an additional phonon~\cite{pringsheim1929zwei, epstein1995observation} was demonstrated experimentally for perovskites micro objects~\cite{ha2016laser}. One of the possible applications of the Purcell factor enhancement in perovskite nanoparticles is more efficient optical cooling at the nanoscale. Indeed, as was predicted theoretically~\cite{tonkaev2019optical}, this effect helps to boost internal QY of PL for MAPbI$_3$ nanoparticles and enhance the absorption of light at pumping wavelength below band gap. As a result, we may expect the cooling down to $\Delta T\approx$--100~K. 

In brief, we anticipate further rapid progress in this research direction driven by the concepts of meta-optics owing to a great potential for applications of halide-perovskite meta-optics in new types of light sources, light-emitting metasurfaces, next-generation displays, quantum signal processing, and efficient lasing. Moreover, meta-optics can have even a broader range of applications highlighting the importance of optically-induced magnetic response, including structural coloring, optical sensing, spatial modulation of light, nonlinear active media, as well as both integrated classical and quantum circuitry and topological photonics, underpinning a new generation of highly-efficient active metadevices.

\begin{acknowledgments}
This work was supported by the Ministry of Education and Science of the Russian Federation (project 14.Y26.31.0010) and the Russian Science Foundation (project 19-73-30023). Y.S.K. acknowledges a support from the Strategic Fund of the Australian National University and also thanks C.~Jagadish for useful comments and his kind suggestion to write this review paper. The authors thank their close collaborators D.~Baranov, K.~Koshelev, I.~Iorsh, A.~Zakhidov E.~Tiguntseva, A.~Furasova, A.~Pushkarev, and T.~Shegai for their valuable input and help. 
\end{acknowledgments}

\bibliographystyle{plain}

%\bibliography{aipsamp}% Produces the bibliography via BibTeX.

\end{document}